\documentclass[12pt]{article}
\usepackage{epsfig}
\usepackage{amsmath}
\usepackage{hhline}
\usepackage{amssymb}
\usepackage{times}
\usepackage{cite}

\newlength{\dinwidth}
\newlength{\dinmargin}
\begin{document}
\newcommand{\pom}{{I\!\!P}}
\newcommand{\reg}{{I\!\!R}}
\newcommand{\slowpi}{\pi_{\mathit{slow}}}
\newcommand{\fiidiii}{F_2^{D(3)}}
\newcommand{\fiidiiiarg}{\fiidiii\,(\beta,\,Q^2,\,x)}
\newcommand{\n}{1.19\pm 0.06 (stat.) \pm0.07 (syst.)}
\newcommand{\nz}{1.30\pm 0.08 (stat.)^{+0.08}_{-0.14} (syst.)}
\newcommand{\fiidiiiful}{F_2^{D(4)}\,(\beta,\,Q^2,\,x,\,t)}
\newcommand{\fiipom}{\tilde F_2^D}
\newcommand{\ALPHA}{1.10\pm0.03 (stat.) \pm0.04 (syst.)}
\newcommand{\ALPHAZ}{1.15\pm0.04 (stat.)^{+0.04}_{-0.07} (syst.)}
\newcommand{\fiipomarg}{\fiipom\,(\beta,\,Q^2)}
\newcommand{\pomflux}{f_{\pom / p}}
\newcommand{\nxpom}{1.19\pm 0.06 (stat.) \pm0.07 (syst.)}
\newcommand {\gapprox}
   {\raisebox{-0.7ex}{$\stackrel {\textstyle>}{\sim}$}}
\newcommand {\lapprox}
   {\raisebox{-0.7ex}{$\stackrel {\textstyle<}{\sim}$}}
\def\gsim{\,\lower.25ex\hbox{$\scriptstyle\sim$}\kern-1.30ex%
\raise 0.55ex\hbox{$\scriptstyle >$}\,}
\def\lsim{\,\lower.25ex\hbox{$\scriptstyle\sim$}\kern-1.30ex%
\raise 0.55ex\hbox{$\scriptstyle <$}\,}
\newcommand{\pomfluxarg}{f_{\pom / p}\,(x_\pom)}
\newcommand{\dsf}{\mbox{$F_2^{D(3)}$}}
\newcommand{\dsfva}{\mbox{$F_2^{D(3)}(\beta,Q^2,x_{I\!\!P})$}}
\newcommand{\dsfvb}{\mbox{$F_2^{D(3)}(\beta,Q^2,x)$}}
\newcommand{\dsfpom}{$F_2^{I\!\!P}$}
\newcommand{\gap}{\stackrel{>}{\sim}}
\newcommand{\lap}{\stackrel{<}{\sim}}
\newcommand{\fem}{$F_2^{em}$}
\newcommand{\tsnmp}{$\tilde{\sigma}_{NC}(e^{\mp})$}
\newcommand{\tsnm}{$\tilde{\sigma}_{NC}(e^-)$}
\newcommand{\tsnp}{$\tilde{\sigma}_{NC}(e^+)$}
\newcommand{\st}{$\star$}
\newcommand{\sst}{$\star \star$}
\newcommand{\ssst}{$\star \star \star$}
\newcommand{\sssst}{$\star \star \star \star$}
\newcommand{\tw}{\theta_W}
\newcommand{\sw}{\sin{\theta_W}}
\newcommand{\cw}{\cos{\theta_W}}
\newcommand{\sww}{\sin^2{\theta_W}}
\newcommand{\cww}{\cos^2{\theta_W}}
\newcommand{\trm}{m_{\perp}}
\newcommand{\trp}{p_{\perp}}
\newcommand{\trmm}{m_{\perp}^2}
\newcommand{\trpp}{p_{\perp}^2}
\newcommand{\alp}{\alpha_s}

\newcommand{\alps}{\alpha_s}
\newcommand{\sqrts}{$\sqrt{s}$}
\newcommand{\LO}{$O(\alpha_s^0)$}
\newcommand{\Oa}{$O(\alpha_s)$}
\newcommand{\Oaa}{$O(\alpha_s^2)$}
\newcommand{\PT}{p_{\perp}}
\newcommand{\JPSI}{J/\psi}
\newcommand{\sh}{\hat{s}}
\newcommand{\uh}{\hat{u}}
\newcommand{\MP}{m_{J/\psi}}
\newcommand{\PO}{I\!\!P}
\newcommand{\xbj}{x}
\newcommand{\xpom}{x_{\PO}}
\newcommand{\ttbs}{\char'134}
\newcommand{\xpomlo}{3\times10^{-4}}
\newcommand{\xpomup}{0.05}
\newcommand{\dgr}{^\circ}
\newcommand{\pbarnt}{\,\mbox{{\rm pb$^{-1}$}}}
\newcommand{\gev}{\,\mbox{GeV}}
\newcommand{\WBoson}{\mbox{$W$}}
\newcommand{\fbarn}{\,\mbox{{\rm fb}}}
\newcommand{\fbarnt}{\,\mbox{{\rm fb$^{-1}$}}}
%
\newcommand{\GeV}{\rm GeV}
\newcommand{\TeV}{\rm TeV}
\newcommand{\pb}{\rm pb}
\newcommand{\cm}{\rm cm}
\newcommand{\hdick}{\noalign{\hrule height1.4pt}}

\def\prp{\perp}
\def\Prp{T}
\def\sx{small-$x$}
\def\kt{\ensuremath{k_\prp}}
\def\kti#1{\ensuremath{k_{\prp #1}}}
\def\pt{\ensuremath{p_\prp}}
\def\pti#1{\ensuremath{p_{\prp #1}}}
\def\qt{\ensuremath{q_\prp}}
\def\qti#1{\ensuremath{q_{\prp #1}}}
\def\xbj{\ensuremath{x_{Bj}}}
\def\xjet{\ensuremath{x_{jet}}}
\def\ptjet{\ensuremath{p_{t\;jet}}}
\newcommand{\dg}{\ensuremath{^{\circ}}}
\def\DJANGO{DJANGO}
\def\RAPGAP{RAPGAP}
\def\CASCADE{CASCADE}
\def\ARIADNE{ARIADNE}
\def\ldcmc{{\small LDCMC}}
\def\PHOJET{PHOJET}
\def\HERACLES{HERACLES}
\def\DISENT{DISENT}
\newcommand{\CCFM}{CCFMa,CCFMb,CCFMc,CCFMd}
\newcommand{\BFKL}{BFKLa,BFKLb,BFKLc}
\newcommand{\LDCMC}{LDCa,LDCb,LDCc,LDCd}
\newcommand{\alphasb}{\alb}
\newcommand{\JETSET}{Jetsetnew}
\newcommand{\LEPTO}{Ingelman_LEPTO65}
\newcommand{\PYTHIAMC}{Jetsetc}
\newcommand{\RAPGAPMC}{RAPGAP206}
\newcommand{\DJANGOMC}{DJANGO}
\newcommand{\HERACLESMC}{HERACLESa,HERACLESb}
\newcommand{\PHOJETMC}{PHOJETa,PHOJETb}
\newcommand{\DGLAP}{DGLAPa,DGLAPb,DGLAPc,DGLAPd}
\newcommand{\qsq}{\ensuremath{Q^2} }
\newcommand{\gevsq}{\ensuremath{\mathrm{GeV}^2} }
\newcommand{\et}{\ensuremath{E_t^*} }
\newcommand{\rap}{\ensuremath{\eta^*} }
\newcommand{\gp}{\ensuremath{\gamma^*}p }
\newcommand{\dsiget}{\ensuremath{{\rm d}\sigma_{ep}/{\rm d}E_t^*} }
\newcommand{\dsigrap}{\ensuremath{{\rm d}\sigma_{ep}/{\rm d}\eta^*} }
\newcommand{\se}[2]{\ensuremath{+ #1 \atop - #2}}

\def\Journal#1#2#3#4{{#1} {\bf #2} (#3) #4}
\def\NCA{\em Nuovo Cimento}
\def\NIM{\em Nucl. Instrum. Methods}
\def\NIMA{{\em Nucl. Instrum. Methods} {\bf A}}
\def\NPB{{\em Nucl. Phys.}   {\bf B}}
\def\PLB{{\em Phys. Lett.}   {\bf B}}
\def\PRL{\em Phys. Rev. Lett.}
\def\PRD{{\em Phys. Rev.}    {\bf D}}
\def\ZPC{{\em Z. Phys.}      {\bf C}}
\def\EJC{{\em Eur. Phys. J.} {\bf C}}
\def\CPC{\em Comp. Phys. Commun.}

\begin{titlepage}

\noindent
\begin{flushleft}
DESY 05-135 \hfill ISSN 0418-9833 \\
August 2005
\end{flushleft}

\vspace{2cm}

\begin{center}
\begin{Large}

{\boldmath \bf  Forward Jet Production in Deep Inelastic Scattering at HERA}

\vspace{2cm}

H1 Collaboration

\end{Large}
\end{center}

\vspace{2cm}

\begin{abstract}
\noindent
The production of forward jets has been measured in deep inelastic
$ep$ collisions at HERA. The results are presented in terms of single
differential cross sections as a function of the Bjorken scaling
variable ($x_{Bj}$) and as triple differential cross sections $d^3
\sigma / dx_{Bj}dQ^2dp_{t,jet}^2$, where $Q^2$ is the four momentum
transfer squared and $p_{t,jet}^2$ is the squared transverse momentum
of the forward jet. Also cross sections for events with a di-jet
system in addition to the forward jet are measured as a function of the
rapidity separation between the forward jet and the two additional
jets. The measurements are compared with next-to-leading order 
QCD calculations and with the predictions of various QCD-based models.
\end{abstract}
\vspace{1.5cm}

\begin{center}
To be submitted to \EJC 
\end{center}

\end{titlepage}

%
%
%
\begin{flushleft}

A.~Aktas$^{10}$,               
V.~Andreev$^{26}$,             
T.~Anthonis$^{4}$,             
S.~Aplin$^{10}$,               
A.~Asmone$^{34}$,              
A.~Astvatsatourov$^{4}$,       
A.~Babaev$^{25}$,              
S.~Backovic$^{31}$,            
J.~B\"ahr$^{39}$,              
A.~Baghdasaryan$^{38}$
% ...

\bigskip{\it
% ...
$ ^{4}$ Inter-University Institute for High Energies ULB-VUB, Brussels;
          Universiteit Antwerpen, Antwerpen; Belgium$^{ c}$ \\
% ...
$ ^{10}$ DESY, Hamburg, Germany \\
% ...
$ ^{25}$ Institute for Theoretical and Experimental Physics,
          Moscow, Russia$^{ l}$ \\
 $ ^{26}$ Lebedev Physical Institute, Moscow, Russia$^{ e}$ \\
% ...
$ ^{31}$ Faculty of Science, University of Montenegro,
          Podgorica, Serbia and Montenegro$^{ e}$ \\
% ...
$ ^{34}$ Dipartimento di Fisica Universit\`a di Roma Tre
          and INFN Roma~3, Roma, Italy \\
% ...
$ ^{38}$ Yerevan Physics Institute, Yerevan, Armenia \\
 $ ^{39}$ DESY, Zeuthen, Germany \\
% ...
$ ^c$ Supported by FNRS-FWO-Vlaanderen, IISN-IIKW and IWT
      and  by Interuniversity
Attraction Poles Programme,
      Belgian Science Policy \\
% ...
$ ^e$ Supported by the Deutsche Forschungsgemeinschaft \\
% ...
$ ^l$ Partly Supported by Russian Foundation
      for Basic Research, grants no. 03-02-17291 and 04-02-16445 \\
}

\end{flushleft}

\newpage
\pagestyle{plain}

\section{Introduction}
The hadronic final state in deep inelastic scattering offers a
rich field of research for QCD phenomena. This includes studies
of hard parton emissions which result in well defined jets,
perturbative effects responsible for multiple gluon emissions and the
non-perturbative hadronisation process.

\begin{figure}[htb]
  \begin{center}
    \vspace*{1mm}
    \vspace*{1cm}
    \epsfig{figure=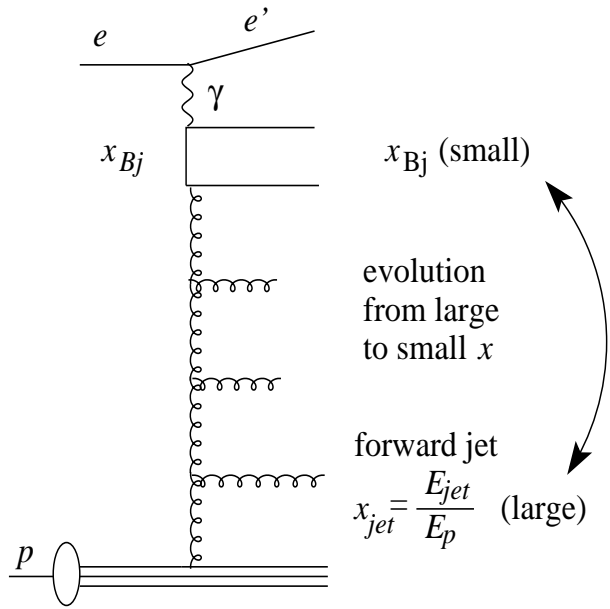,width=8cm,height=8cm}
    \caption{{\it
Schematic diagram of $ep$ scattering with a forward jet taking a
fraction $x_{jet}=E_{jet}/E_p$ of the proton momentum. The
evolution in the longitudinal momentum fraction, $x$, 
from large $\xjet$ to small $\xbj$ is indicated. 
    \label{forward_jet_fig}}}
  \end{center}
\end{figure}

HERA has extended the available region in the Bjorken  scaling
variable, $\xbj$, down to values  of $\xbj \simeq 10^{-4} $, for
values of the four momentum transfer squared, $Q^2$, larger than a
few GeV$^2$, where perturbative calculations in  QCD are expected to
be valid. At these low $\xbj$ values, a parton in the proton can
induce a QCD cascade, consisting of several subsequent parton
emissions, before eventually an interaction with  the virtual photon
takes place (Fig.~\ref{forward_jet_fig}). QCD calculations based on
``direct" interactions between a point-like photon and a parton from
an evolution chain, as given by the DGLAP scheme~\cite{Gribov:1972rt,
Gribov:1972ri, Lipatov:1974qm, Altarelli:1977zs, Dokshitzer:1977sg},
are successful in reproducing the strong rise of $F_2(\xbj,Q^2)$ with
decreasing $\xbj$ over a large $Q^2$  range~\cite{Adloff:1999ah,
Adloff:2000qk,Chekanov:2001qu,Chekanov:2003yv}. The DGLAP evolution
resums leading $\log(Q^2)$ terms. This approximation, however, may
become inadequate for small $\xbj$, where $\log(1/x)$ terms become
important in the evolution equation. In this region the BFKL
scheme~\cite{Kuraev:1976ge, Kuraev:1977fs, Balitsky:1978ic} is
expected to describe the data better, since this evolution equation
sums up terms in $\log(1/x)$.

Significant deviations from the simple leading order (LO) DGLAP
approach are observed in the fractional rate of di-jet
events~\cite{Adloff:1998vc, Adloff:1998st, Chekanov:2001fw},
inclusive jet production~\cite{Adloff:1997nd, Adloff:2002ew},
transverse energy flow~\cite{Aid:1995we, Adloff:1999ws} and $p_t$
spectra of charged particles~\cite{Adloff:1996dy}. Extending the
calculations from LO to next-to-leading order (NLO) accuracy accounts for some of the
deviations observed in jet production, but at low $\xbj$ and
low $Q^2$ the description of the
measurements is still unsatisfactory. Next-to-next-to-leading order (NNLO) calculations do not exist
so far and therefore higher order contributions can only be approximated
by phenomenological QCD models, based on LO matrix element
calculations together with parton shower evolution. Ascribing
partonic structure to the virtual photon and thus considering so
called resolved photon processes, including parton showers from both
the photon and the proton side, results in an improved description of
the data including particle production in the forward region (the
angular region close to the proton beam
direction)~\cite{Adloff:1998vc, Breitweg:2000sv, Jung:1998fu, Jung:1999eb,
kamildijet, Adloff:1999zx, Aktas:2004rb}. The colour dipole model
(CDM)~\cite{Andersson:1988gp, Lonnblad:1994wk}, which assumes gluon
emissions to originate from independently radiating colour dipoles,
is in fairly good agreement with the measurements. This suggests
that different parton dynamics, not included in the DGLAP
approximation, are responsible for the observed deviations.

The large phase space available at low $x_{Bj}$ makes the production
of forward jets a particularly interesting process for the study of
parton dynamics, since jets emitted close to the proton direction 
lie well away in rapidity from the photon end of the
evolution ladder (Fig.~\ref{forward_jet_fig}). Here a new measurement
of forward jet production is presented using data collected in 1997
with the H1 detector, comprising an
integrated luminosity of 13.7 pb$^{-1}$. The enlarged statistics
allows to study more differential distributions than previously
presented~\cite{Adloff:1998fa, Breitweg:1998ed,
Chekanov:2005yb}. The proton energy is 820~GeV and the positron energy is
27.6~GeV which correspond to a centre-of-mass-energy of
$\sqrt{s}\approx  300$~GeV.

Measurements are presented in regions of phase  space
where the DGLAP approximation might be insufficient to describe the
parton dynamics. In inclusive forward jet production this is expected
to be the case when the transverse  momentum squared of the jet and
the photon virtuality are of similar order. More exclusive final
states, like those containing a  di-jet system in addition to the forward
jet (called `2+forward jet'), provide a further handle to control
the parton dynamics.   The forward jet measurements are compared to
LO and  NLO di- and three-jet calculations,
and different phenomenological QCD models. 
This measurement is complementary to a
similar measurement of $\pi^o$-production in the forward direction, 
which has been presented
in~\cite{Aktas:2004rb}. 

\section{QCD Models and Theoretical Calculations}
\label{sec: mc}
The conventional description of the parton cascade is
given by the DGLAP evolution equations. The basic assumption is that
the leading contribution comes from cascades with strong ordering in the
virtualities of the parton propagators in the evolution chain, with
the largest virtualities reached in the hard scattering with the
photon. This implies strong ordering of the transverse momenta of the
emitted partons ($k_t$). Since their virtualities and transverse
momenta squared are small compared to the hard scale, $Q^2$, the propagators
can be treated as massless and assumed to be collinear with the
incoming proton (collinear approach). The interaction is assumed to
take place with a point-like photon (DGLAP direct).

If the scale of the hard subprocess is larger than $Q^2$, the
structure of the virtual photon might be resolved and the
interaction take place with one of the partons in the photon. In
this case a partonic structure is assigned to the photon and a
photon parton density function is convoluted with the matrix
element, which within the DGLAP model means that two evolution
ladders are introduced, one from the photon side and one from the
proton side of the hard subprocess. This is called the resolved
photon model~(DGLAP resolved) and is described in~\cite{Jung:1998fu,Jung:1999eb}.

The BFKL ansatz predicts
strong ordering in the longitudinal momentum
fraction of the parton propagators but no ordering in their
virtualities. This means that the virtualities and the transverse
momenta of the propagators can  take any kinematically allowed value
at each splitting. One consequence of this is that the matrix element
must be taken off mass-shell and convoluted with parton distributions
which take the transverse momenta of the propagators into account
(unintegrated parton densities).

The CCFM equation~\cite{Ciafaloni:1987ur, Catani:1989yc, Catani:1989sg, Marchesini:1994wr} 
provides a bridge between the DGLAP and
BFKL descriptions by resumming both $\log (Q^2)$ and $\log
(1/x)$ terms in the relevant limits, and is expected to be valid in a wider $x$ range. The CCFM
equation leads to parton emissions ordered in angle. An unintegrated
gluon density is used as input to calculations based on this model.

A different approach to the parton evolution is given by the 
colour dipole model (CDM),
in which the emissions are generated by colour dipoles,
spanned between the partons in the cascade. Since the dipoles
radiate independently there is no ordering in the transverse
momenta of the emissions and the behaviour of the parton showers is
in that sense similar to that in the BFKL case.

The measurements performed here are compared to several QCD models:

\begin{itemize}
\item The \RAPGAP~\cite{Jung:1993gf} Monte Carlo program, which
uses LO matrix elements supplemented with initial and final state
parton showers generated according to the DGLAP evolution scheme for
the description of DIS processes (RG-DIR). It can be interfaced
to \HERACLES~\cite{Kwiatkowski:1990es}, which
simulates QED-radiative effects. \RAPGAP\ also offers the
possibility to include contributions from processes with resolved
transverse virtual photons (RG-DIR+RES).

\item The \DJANGO~\cite{Charchula:1994kf}
program with the CDM as implemented in
\ARIADNE~\cite{Lonnblad:1992tz}. Parameters of \ARIADNE~are tuned
using the CTEQ6M~\cite{Pumplin:2002vw} parton density functions and
the data sets~\cite{Adloff:1996dy, Adloff:2000tq, Adloff:1999ws}.

\item The \CASCADE\ Monte Carlo program~\cite{Jung:2000hk,
Jung:2001hx}, which is based on the CCFM formalism~\cite{Ciafaloni:1987ur,
Catani:1989yc, Catani:1989sg, Marchesini:1994wr}. Two different
versions of the unintegrated gluon density are used, J2003-set-1 and
set-2~\cite{Hansson:2003xz}. The difference between these two sets is
that in set-1 only singular terms are included in the splitting
function, whereas  set-2 also takes the non-singular terms into
account. These unintegrated gluon densities are determined from fits to
the $F_2(x,Q^2)$ data obtained by H1 and ZEUS in 1994 and 1996/97.

\end{itemize}

Simulated events from the \RAPGAP\ (RG-DIR) and \DJANGO\ Monte
Carlo programs are processed through the detailed H1
detector simulation~\cite{Brun:1978fy}
in order to test the understanding of the
detector and to extract correction factors.

The forward jet cross sections are compared to LO ($\alpha_s$) and NLO
($\alpha_s^2$) calculations of di-jet production via direct photon interactions as obtained
from the DISENT program~\cite{Catani:1996jh, Catani:1996vz} . Since the 
jet search is performed in the Breit frame the selected events always 
contain at least one jet in addition to the forward jet, such that 
comparisons with the DISENT predictions are adequate. The renormalisation scale
$(\mu_r^2)$ is given by the average $p_t^2$ of the di-jets from the
matrix element~($\langle p_{t,\textrm{di-jets}}^2 \rangle$), while
the factorisation scale $(\mu_f^2)$ is given by the average $p_t^2$
of all forward jets in the selected sample\footnote{For the triple
differential forward jet cross section, $d^3 \sigma /
dx_{Bj}dQ^2dp_{t,jet}^2$, this means different factorisation scales
for the three different $p_{t,jet}$ bins.} ($\langle
p_{t,jet}^2 \rangle$). The calculations are corrected for
hadronisation effects, which are estimated using CASCADE together
with the KMR parton density function~\cite{Kimber:2001sc}. The KMR
parton density function takes only the matrix element and one
additional emission into account and should therefore be suitable for
correcting the NLO di-jet calculations. The correction factors for
hadronisation effects $(1+\delta_{\textrm{HAD}})$ are determined by
calculating the ratio bin-wise between the hadron and parton level
cross sections, obtained using the same jet algorithm and
kinematic restrictions.

In the analysis of events with two jets in addition to the
forward jet, the measured cross sections are compared to the predictions of
NLOJET++~\cite{Nagy:2001xb}. This program provides perturbative calculations of cross
sections for three-jet production in DIS at NLO ($\alpha^3_s$) accuracy. 
In \linebreak NLOJET++, where the factorisation scale can be defined for each event,
$\mu_r^2$ and $\mu_f^2$ are set to the average $p_t^2$ of the
forward jet and the two hardest jets in the event.
The
NLOJET++ calculations are corrected to hadron level using
CASCADE together with the unintegrated gluon density~J2003
set-2~\cite{Hansson:2003xz}.

The NLO calculations by DISENT and NLOJET++ are performed using the
CTEQ6M~\cite{Pumplin:2002vw} parametrisation of the parton distributions in the proton.
The uncertainty in the NLO calculations originating from the PDF
uncertainty is estimated by using the CTEQ eigenvector sets
according to~\cite{Pumplin:2002vw}.
The scale
uncertainty for these calculations is estimated by simultaneously
changing the renormalisation and factorisation scales
($\mu_r^2,\mu_f^2$) by a factor of 4 up and 1/4 down.
In CASCADE the renormalisation scale $(\mu_r^2)$ is changed by the same
factors and in each case the unintegrated gluon density is adjusted
such that the prediction of CASCADE describes the inclusive $F_2$
data~\cite{Jung:2004gs, Jung:2004bq}.
The forward jet cross section is then calculated to estimate the upper and
lower limit of the scale uncertainty. The resulting
uncertainty in the cross section prediction is less
than 10\% at the smallest $x_{Bj}$ and decreases for
higher $x_{Bj}$ (these errors are not shown in the figures).
The
parton densities and the scales used in the
QCD calculations are given in table~\ref{generatortable}.

\begin{table}[h]
\begin{center}
\scalebox{0.83}{\begin{tabular}{|c||c|c|c|c|}
\hline
 & \CASCADE\ & RG-DIR/RES & DISENT & NLOJET++ \\
\hline
\hline
$\mu_r^2$ & $m^2+\langle p_{t,\textrm{di-jets}}^2 \rangle$ & $Q^2+\langle p_{t,\textrm{di-jets}}^2 \rangle$ & $\langle p_{t,\textrm{di-jets}}^2 \rangle$&$(p_{t,jet1}^2+p_{t,jet2}^2+p_{t,\textrm{fwdjet}}^2)/3$ \\
\hline 
$\mu_f^2$ & $\hat{s}+Q^2$ & $Q^2+ \langle p_{t,\textrm{di-jets}}^2 \rangle$ & $\langle p_{t,jet}^2 \rangle$&$(p_{t,jet1}^2+p_{t,jet2}^2+p_{t,\textrm{fwdjet}}^2)/3$\\
\hline
proton PDF  & J2003 set-1 $\&$-2 & CTEQ6L~\cite{Pumplin:2002vw} & CTEQ6M & CTEQ6M\\
\hline
photon PDF  & - &  SaS1D~\cite{Schuler:1996fc} (RES only) & - & -\\
\hline
\end{tabular}}
\end{center}
\caption{{\it The renormalisation ($\mu_r^2$) and factorisation ($\mu_f^2$)
scales, and the parton density functions used in the
different programs. The average squared transverse momentum of
the forward jet, $\langle p_{t,jet}^2  \rangle$, is 45 GeV$^2$ for the single
differential
forward jet cross section, and 24, 55 and 183 GeV$^2$ for the
three different $p^2_t$-bins in the triple differential cross
sections. $\hat{s}$ is the invariant mass squared of the di-quark system.
\label{generatortable}}}
\end{table}

In~\cite{Kramer:1999jr}
next-to-leading order calculations of the forward jet cross section
are presented, in which the contributions from direct and
resolved virtual photons are taken into account in a consistent
way. The
inclusion of NLO contributions from the resolved part corresponds to
an additional gluon emission in a direct process and thus may
constitute an approximation of the NNLO
direct cross section. 

\section{The H1 Detector} A detailed description of the H1 detector
can be found in~\cite{Abt:1996xv, Abt:1996hi, Appuhn:1996na}.  The
detector elements important for this analysis are described below.
The coordinate system of H1 is
defined such that the positive $z$ axis is in the direction of the
incident proton beam. The polar angles are
defined with respect to the proton beam direction.

The interaction vertex is determined with the central tracking
detector consisting of two concentric drift chambers (CJC) and two
concentric $z$ drift chambers (CIZ and COZ). The kinematic variables
$x$ and $Q^2$ are determined from a measurement of the scattered
electron in the lead-scintillating fibre calorimeter (SpaCal) and the
backward drift chamber (BDC), covering the polar angular range $153\dg <
\theta < 177\dg$. 

The SpaCal has an electromagnetic section with an energy resolution
of 7\%$/\sqrt{E/GeV} \oplus 1$\%, which together with a hadronic section
represents a total of two interaction lengths. Identification of the
scattered electron is improved using the BDC, situated in front of
the SpaCal. The scattering angle of the electron is determined from
the measured impact position in the BDC and the reconstructed primary
interaction vertex.

The hadronic final state is reconstructed with the Liquid Argon
calorimeter (LAr), the central tracking detector and the SpaCal. The LAr
calorimeter is of a sandwich type with liquid argon as the active
material. It covers the range $4\dg < \theta < 154\dg$. In test beam 
measurements pion induced hadronic energies were reconstructed
with a resolution of about 50\%$/\sqrt{E/GeV} \oplus 2$\%~\cite{Andrieu:1993tz}.
The measurement of charged particle momenta provided by the central
tracking detector is performed in a solenoidal magnetic field of 1.15
T with a precision of $\sigma_p/p^2 = 0.003$ GeV$^{-1}$. 

The luminosity is determined from the rate of Bethe-Heitler events
($e+p\rightarrow e+\gamma+p$) with a precision of 1.5\%.

The scattered electron is triggered by its energy deposition in
the SpaCal. For events used in this analysis, with the electron energy
required to be above 10 GeV, the trigger efficiency is essentially
100\%.

\section{Experimental Strategy and Phase Space Definition}
\label{sec:phasespace} Differences between the various
approaches to the modelling of the parton cascade dynamics are most
prominent in the region close to the proton remnant direction,
i.e. away from the photon side of the ladder. This can be
understood from the fact that the strong ordering in virtuality of
the DGLAP description gives the softest $k_t$-emissions closest to the
proton whereas in the BFKL model the emissions can be arbitrarily
hard in this region, as long as they are kinematically allowed.

In most of the
HERA kinematic range the DGLAP approximation is valid. A method to
suppress contributions from DGLAP like events is to select events
with a jet close to the proton direction (a forward jet) with the
additional constraint that the squared transverse momentum of this
jet, $p_{t,jet}^2$, is approximately equal to the virtuality of the
photon propagator, $Q^2$  (see
Fig.~\ref{forward_jet_fig}). This will suppress contributions with
strong ordering in virtuality as is the case in DGLAP evolution. If,
at the same time, the forward jet is required to take a large
fraction of the proton momentum, $x_{jet} \equiv E_{jet}/E_p$, such
that $x_{jet} \gg \xbj$, the phase space for an evolution with
ordering in the longitudinal momentum fraction, as described by BFKL,
is favoured. By requiring $x_{jet} \gg \xbj$ contributions from
zeroth order processes are also suppressed.
Based on calculations in the leading log approximation of the BFKL
kernel, the cross section for  DIS events at low $\xbj$ and large
$Q^2$ with a forward jet \cite{Mueller:1990er, Mueller:1990gy} is
expected to rise more rapidly with decreasing $\xbj$ than expected
from DGLAP based calculations.

DIS events are selected by requiring a scattered electron in the
backward SpaCal calorimeter and a matching track in the backward drift
chamber (BDC), applying the following cuts:

\begin{center}
\begin{tabular}{c}
$E'_{e} > 10  \mbox{ GeV}$\\
$ 156 \dg < \theta_e < 175\dg$\\
$ 0.1 < y < 0.7$ \\
$ 0.0001 < x_{Bj} < 0.004$ \\
$ 5\mbox{ GeV}^2 < Q^2 < 85 \mbox{ GeV}^2$ 
\label{discuts}
\end{tabular}
\end{center}
where $E'_e$ is the energy of the scattered electron, $\theta_e$ its polar angle, 
and $y$ is the inelasticity variable. The lower cut on $Q^2$ and and the upper on $y$ reduce the
background from photoproduction.

Jets are defined using the $k_t$-jet algorithm~\cite{Catani:1993hr,
Catani:1992zp} with combined calorimeter and track
information~\cite{Adloff:1997mi} as input (applied in the
Breit-frame). 
The selection further requires the reconstruction of at least 
one jet in the laboratory frame, satisfying the cuts below:

\begin{center}
\begin{tabular}{c}
$p_{t,jet} > 3.5 $ GeV \\
$7\dg < \theta_{jet} < 20\dg $  \\
$ \xjet > 0.035 $\\
\label{fjcuts}
\end{tabular}
\end{center}
where the $p_{t,jet}$- and $\theta_{jet}$-cuts are applied in the
laboratory frame.
If there is more than one jet fulfilling these requirements the most
forward is chosen. 
For the single differential cross section measurement
an additional cut $ 0.5 < r = p^2_{t,jet}/Q^2 < 5 $ is applied. 

Data are presented as single differential cross-sections as a
function of $\xbj$ ($d\sigma/d\xbj$), and triple differential
cross-sections as a function of $\xbj$ in bins of $Q^2$ and
$p^2_{t,jet}$ \linebreak ($d^3 \sigma / dx_{Bj}dQ^2dp_{t,jet}^2$). Another event
sample, called the `2+forward jet' sample, is selected by requiring
that, in addition to the forward jet, at least two more jets are found.
Out of these, the two with the highest transverse momenta are chosen. 
This provides further constraints on the
kinematics at the expense of reducing the data sample. 

For the `2+forward jet' sample the $p_t$ is required to be larger than
6~GeV for all 3 jets. The other cuts on the forward jet are kept the
same as specified above, and no $p^2_{t,jet}/Q^2$-cut is applied.
The two additional jets are required to lie 
in pseudorapidity, $\eta =-\ln \tan(\theta/2)$, between the electron and the
forward jet, $\eta_e < \eta_{jet1} < \eta_{jet2} <
\eta_{fwdjet}$.

The final numbers of events used for the single and the triple
differential forward jet cross section are 17316 and 23992,
respectively. The number of selected `2+forward jet' events is 854.

\section{Correction Factors and Systematic Uncertainties} 
\begin{figure}[htb]
\hspace{1.5cm} {\bf{Forward jets}}
  \begin{center}
    \epsfig{figure=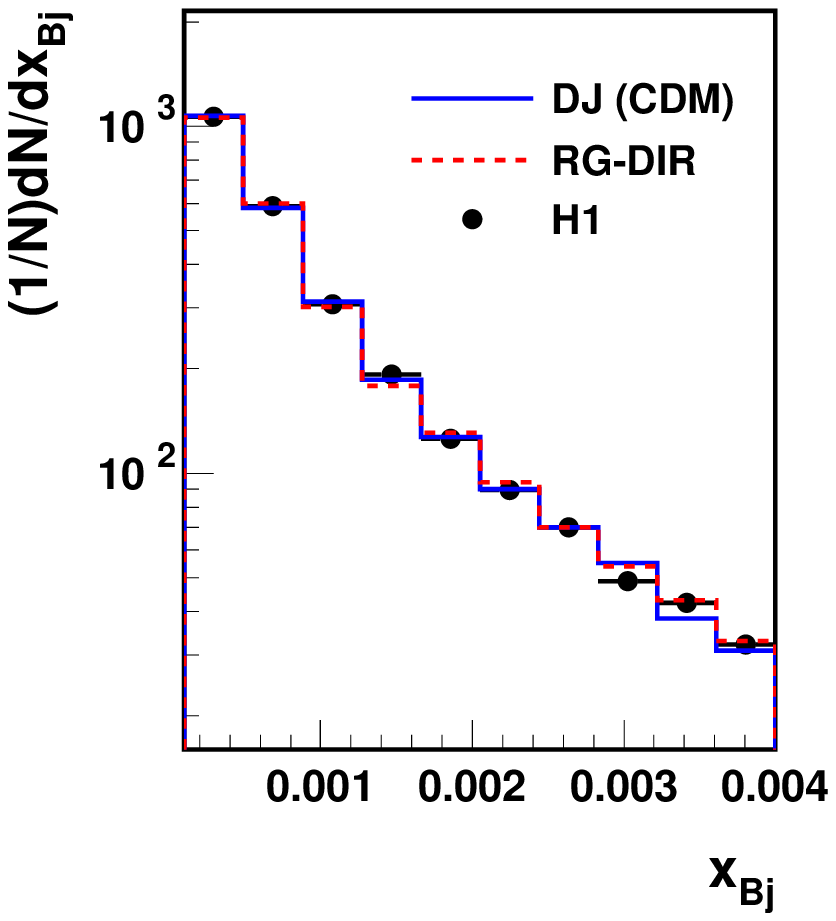,height=5.5cm}
    \epsfig{figure=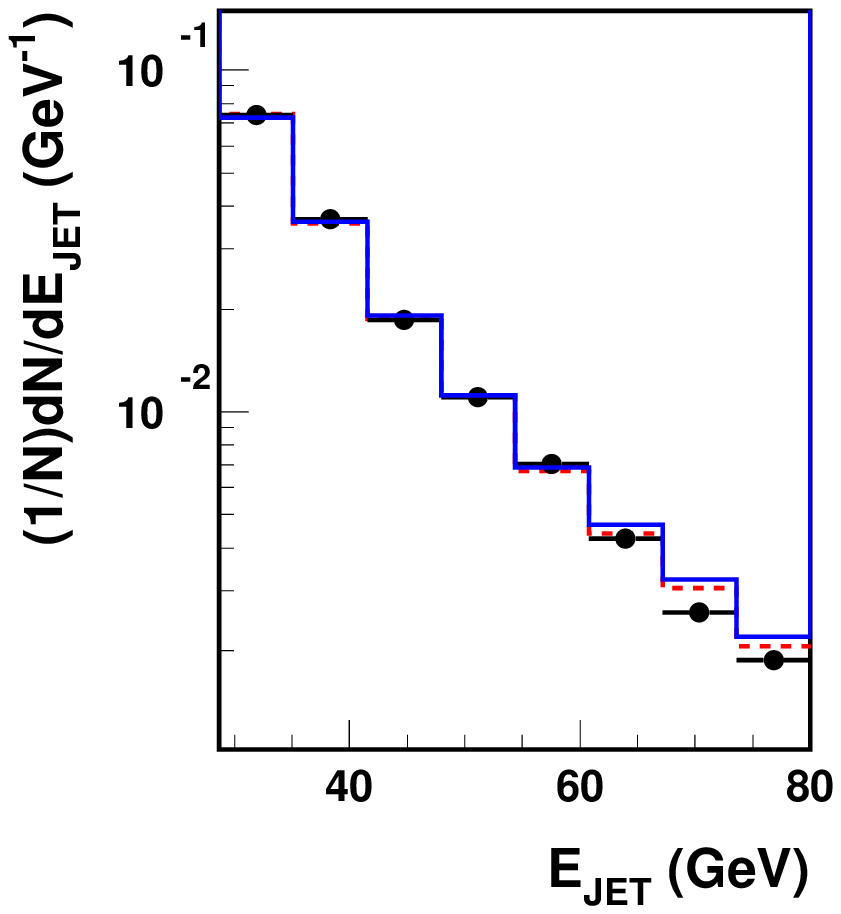,height=5.5cm}
    \epsfig{figure=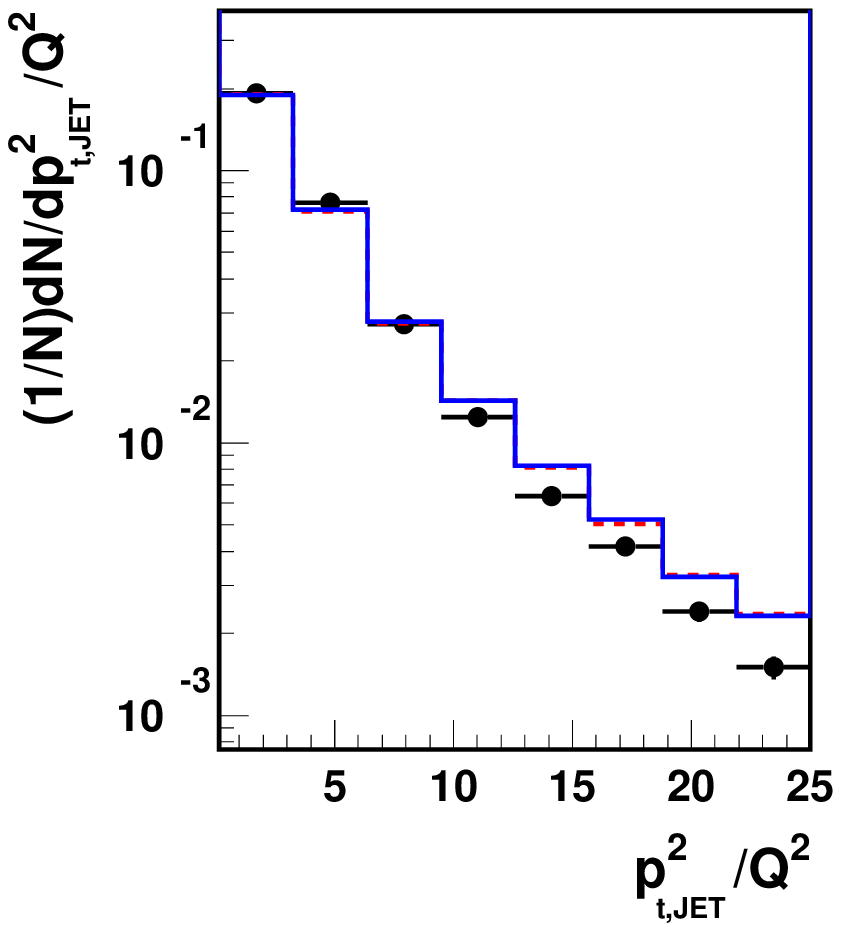,height=5.5cm}\\
\end{center}
\hspace{1.5cm} {\bf{Forward jets with} {\boldmath{$0.5 < p^2_{t,jet}/Q^2 < 5$}}}
\begin{center}
    \epsfig{figure=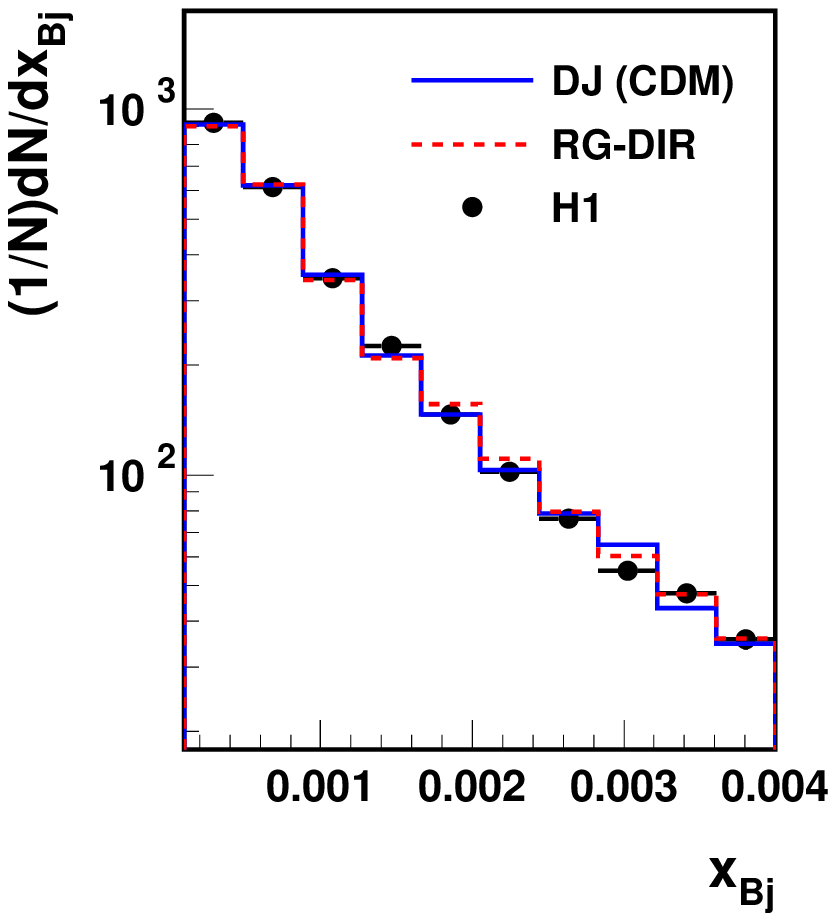,height=5.5cm}
    \epsfig{figure=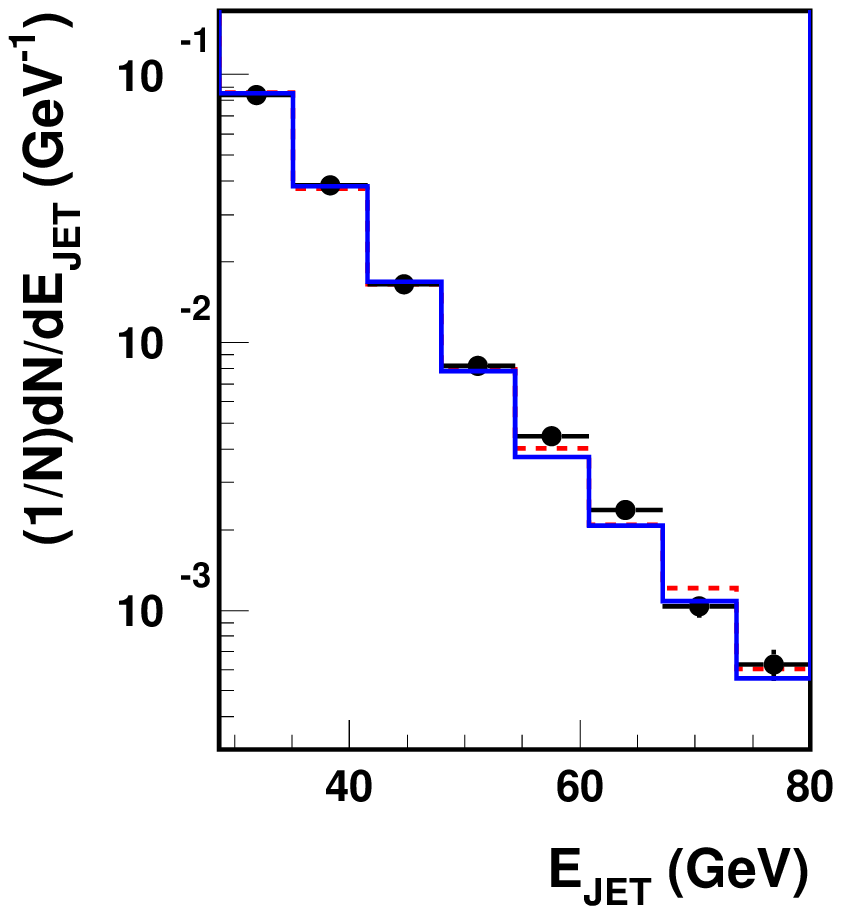,height=5.5cm}
    \epsfig{figure=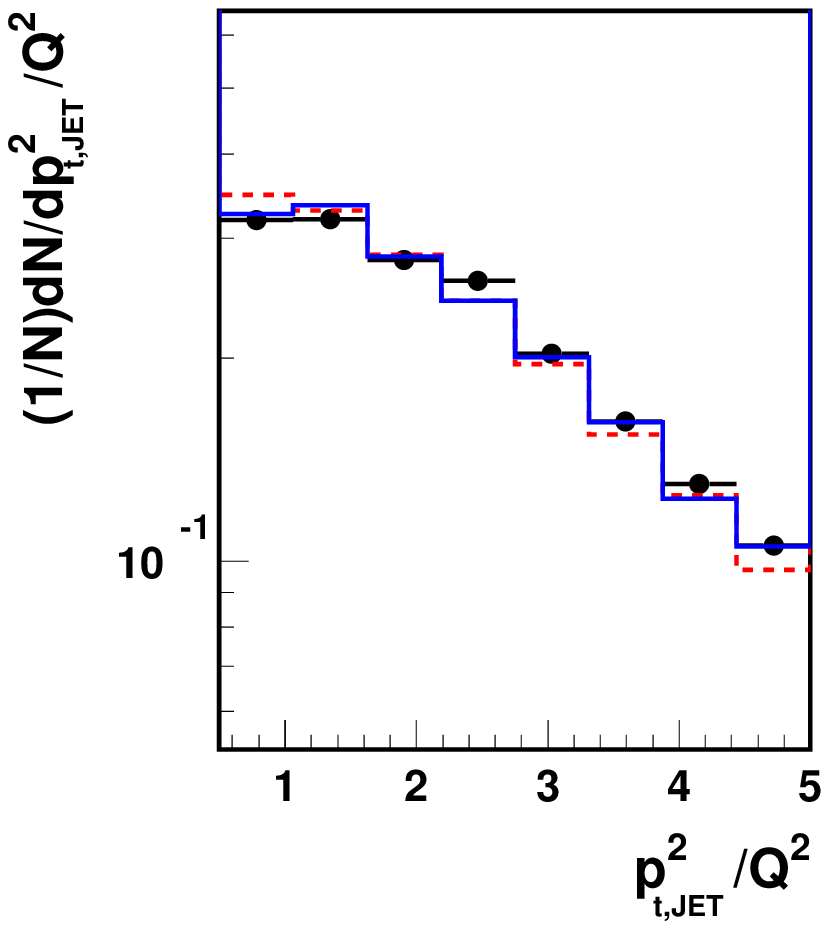,height=5.5cm}\\
    \caption{{\it
    Control plots for the forward jet selection. The sample
    with no $p_{t,jet}^2/Q^2$-cut applied (upper) and the sample with the $0.5 < p_{t,jet}^2/Q^2 < 5$-cut
    applied (lower) are shown. The distributions are at detector level and normalised to unity. All
    variables are measured in the laboratory frame. Comparisons are made to the predictions of
    the DJANGO (full line) and RAPGAP (dashed line) Monte Carlo programs.
    \label{fj_fjchecks}}}
  \end{center}
\end{figure}

The
\RAPGAP\ and \DJANGO\ programs, together with a simulation of the H1
detector, are used to correct the data for acceptances,
inefficiencies, and bin to bin migrations due to the finite detector
resolutions. The shapes of the distributions of the DIS kinematic
variables and the jet variables for the forward jet sample, as
defined in section~\ref{sec:phasespace}, are compared to the
predictions from \RAPGAP\ and DJANGO. This is done by reweighting
the Monte Carlo $\xbj$ distributions to give the best possible
agreement with data and by studying how well the distributions of the
other kinematic variables are described. The distributions are
reproduced equally well by the predictions of \RAPGAP\ and
\DJANGO\ after the detector simulation. In
Fig.~\ref{fj_fjchecks} detector level distributions are shown for
$x_{Bj}$, $E_{jet}$ and $p^2_{t,jet}/Q^2$ for the forward jet samples, with
and without the $p^2_{t,jet}/Q^2$-cut applied. These distributions are
normalised to the number of events and thus give a shape comparison
to investigate the understanding of the detector, independently of
the normalisation of the physics models.

The hadron level cross sections are extracted by applying correction
factors to the data in order to take
detector effects into account. The correction factors are calculated
as the ratio of the CDM Monte Carlo prediction at the hadron and
detector levels, in a bin-by-bin procedure. These factors correct the
data from detector level to non-radiative hadron level, i.e.~the data
are also corrected for QED radiative effects. RAPGAP and CDM give
similar values over the full kinematic range covered in this
investigation. The correction factors are generally between 0.7 and 1.2 but in
a few kinematic bins they reach values of 0.5 or 1.4 due to limited
resolution of the jet quantities. The variations in the correction
factors between the two Monte Carlo models are included in the
systematic error.  

The purity and acceptance\footnote{The purity (acceptance) is
obtained from the same Monte Carlo simulations as used for the
correction factors and is defined as the number of simulated events which
originate from a bin and are reconstructed in it divided by the number of
reconstructed (generated) events in that bin.} 
are found to be larger than 30\% in all bins. For
the `2+forward jet' analysis they are larger than 40\% in all bins.

The systematic errors are estimated for each data point separately
as the quadratic sum of the individual errors described below. 
The following systematic errors are
considered:

\begin{itemize}

\item The hadronic energy scale uncertainty is determined to be 4\%.
In order to estimate the related uncertainty of the measured forward
jet cross section, the reconstructed hadronic energies in the
DJANGO/ARIADNE simulation were increased and decreased by this amount.
The average resulting error is typically 8\% for both the single
differential forward jet cross section and the triple differential
forward jet cross section, and 13\% for the `2+forward jet' cross
section.

\item The electromagnetic energy scale as measured in the SpaCal is
known to an accuracy of 1\%. Changing the scale by this amount in the
forward jet cross section calculations using DJANGO/ARIADNE results
in an average systematic error of typically 3\% for the single and
triple differential measurement, and 1\% for the `2+forward jet'
measurement.

\item The uncertainty on the measured scattering angle of the
electron is estimated to be \linebreak 1 mrad, which contributes typically 1\%
to the error in the forward and `2+forward jet' cross section.

\item The error from the model dependence is taken as the
difference between the correction factors calculated from the
DJANGO/ARIADNE and the RG-DIR Monte Carlo programs. Taking this
variation into account yields a systematic error of about 5\% for
the single differential forward jet cross section, 8\% for the triple differential
case and 13\% in the `2+forward jet' cross section.

\item  The PHOJET~\cite{Engel:1995yd, Engel:1994vs} Monte Carlo generator was used in
order to estimate the extent to which DIS forward jet events could be
faked by photoproduction ($Q^2 \sim 0$ GeV$^2$) background. The
influence on the measurement is found to be negligible. The error
attributed to this source of uncertainty is taken to be 1\%.

\item The uncertainty of the luminosity measurement is
estimated to be 1.5\%.

\end{itemize}

The averages of these sums are 10\%, 12\% and 14\% for the single
differential, triple differential and the `2+forward jet' cross
section, respectively. In the figures the systematic errors due to 
the energy scale uncertainty of the calorimeters ($\Delta_{\textrm{Syst1}}$) 
are shown separately as bands around the data points, whereas the 
other systematic errors ($\Delta_{\textrm{Syst2}}$) are included 
in the error bars together with the statistical errors. The 
errors are given separately in the tables.

\section{Results}
\label{sec:results}
\subsection{Single Differential Cross Section}
\label{sec:inclusive} The measurement of the single differential
forward jet cross section is presented at the hadron level in the
phase space region defined in section~\ref{sec:phasespace}. The
phase space for DGLAP evolution is suppressed by the  
additional requirement $0.5 < p^2_{t,jet}/Q^2 < 5$ as discussed 
in section~\ref{sec:phasespace}.

\begin{figure}[htb]
  \begin{center}
    \vspace*{1mm}
    \vspace*{1cm}
     \epsfig{figure=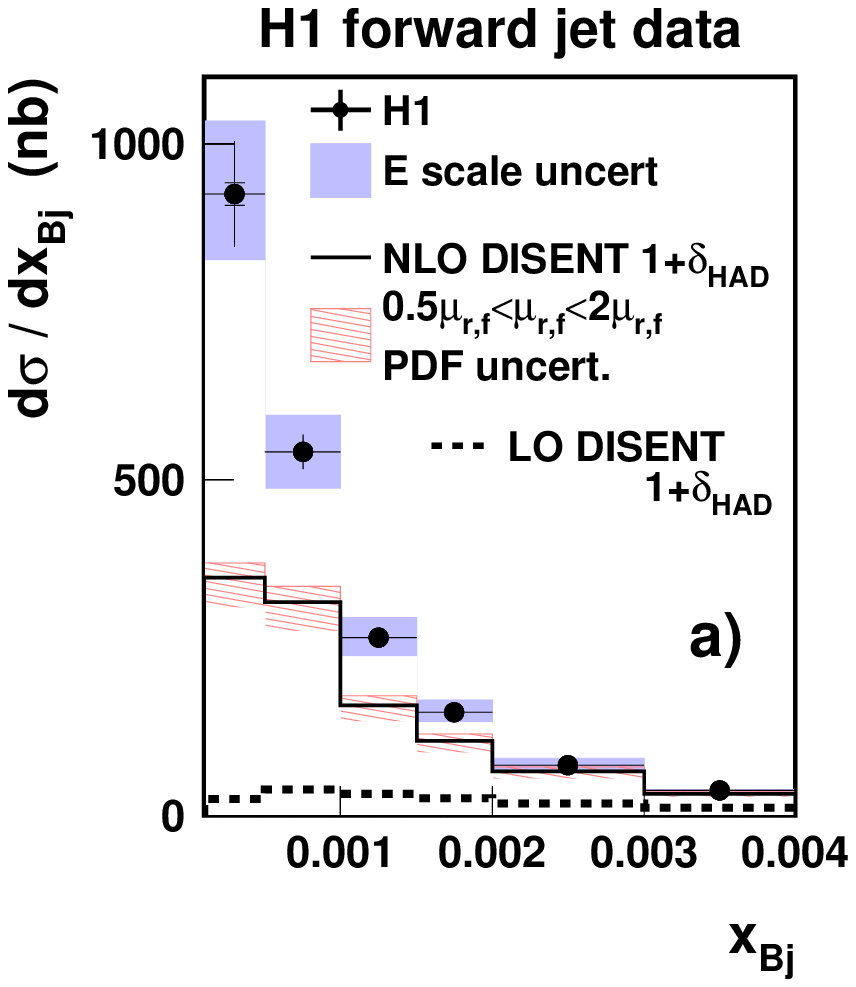,height=5.5cm}
     \epsfig{figure=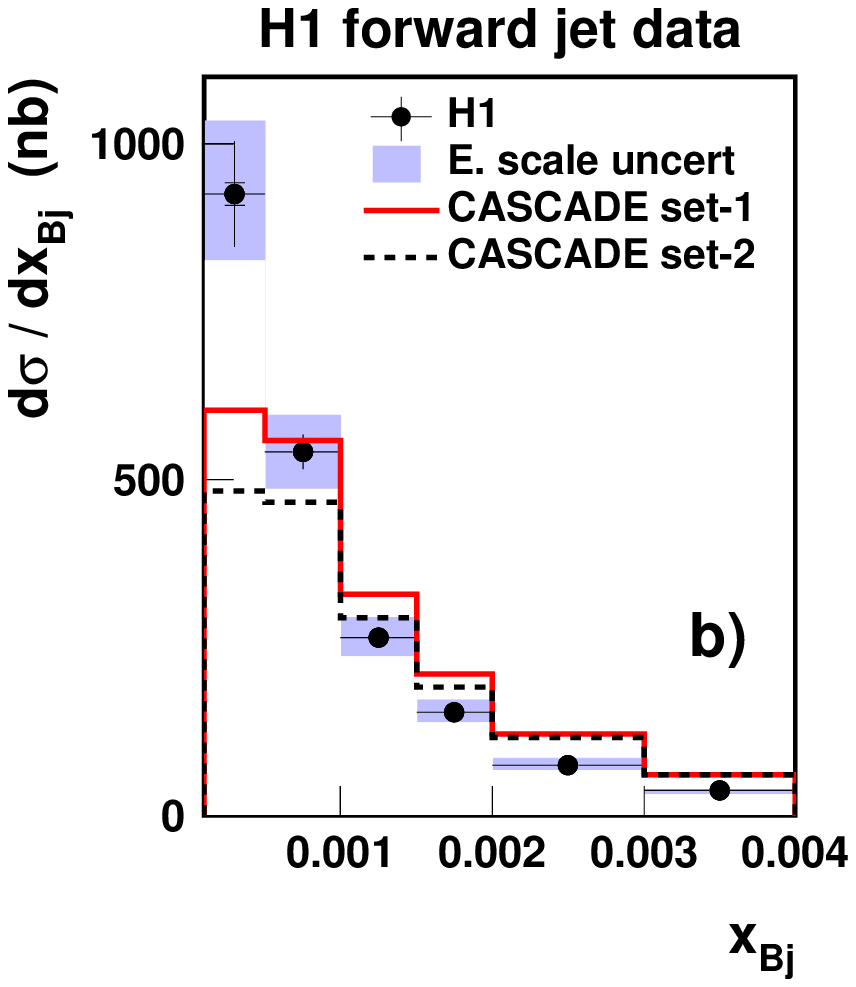,height=5.5cm}
     \epsfig{figure=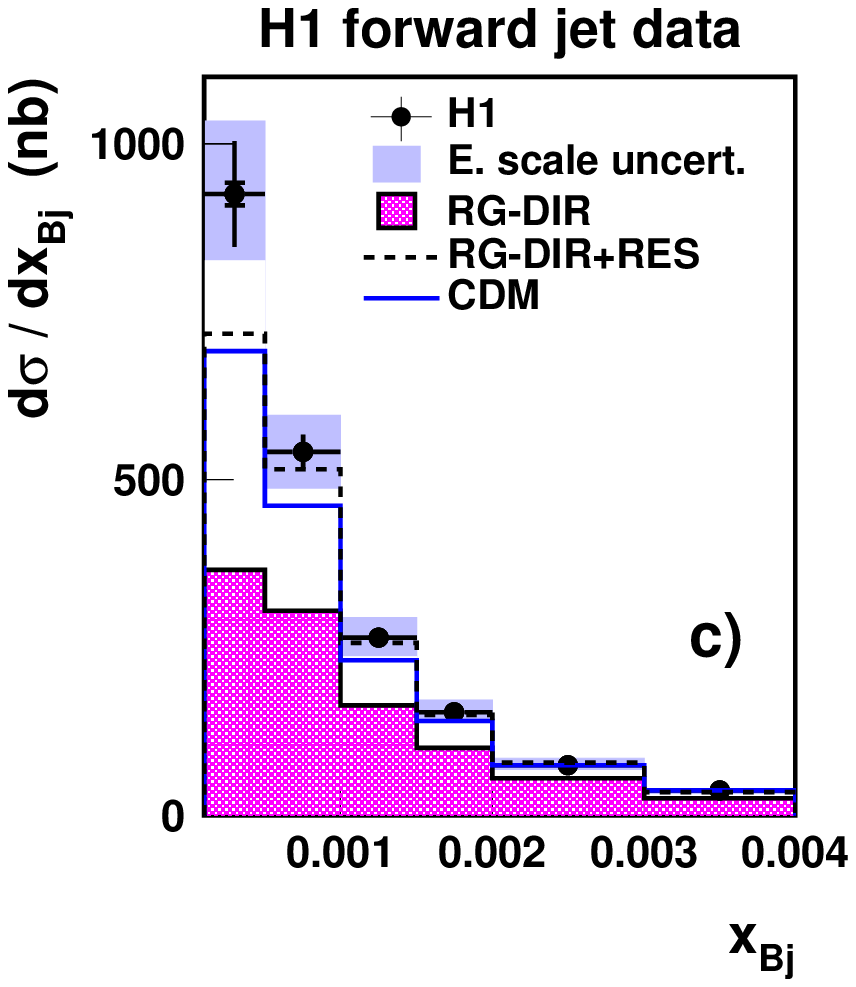,height=5.5cm}
   \caption{{\it
    The hadron level cross section for forward jet production as a function of
$\xbj$ compared to NLO predictions from DISENT
(a) and to QCD Monte Carlo models (b and c). The shaded band
around the data points shows the error from the uncertainties
in the calorimetric energy scales. The hatched band around the NLO
calculations illustrates the theoretical uncertainties in the
calculations, estimated as described in the text. The dashed line
in (a) shows the LO contribution.
    \label{xfj_djcorr}}}
  \end{center}
\end{figure}

The measured single differential forward jet cross sections are
listed in table~\ref{tblxsecxfj}. In Fig.~\ref{xfj_djcorr}a they
are compared with LO ($\alpha_s$) and NLO ($\alpha^2_s$) calculations from DISENT. The
calculations are multiplied by $(1+\delta_{\textrm{HAD}})$ to correct
to the hadron level. The uncertainty from the factorisation and
renormalisation scales, and the uncertainty in the PDF
parametrisation, are added in quadrature to give the total
theoretical error, which is shown as a band around the histogram
presenting the theoretical prediction. 
In Fig.~\ref{xfj_djcorr}b and c the data are compared to the
various QCD models.

In Fig.~\ref{xfj_djcorr}a it can be observed that, at small $x_{Bj}$,
the NLO di-jet calculations from DISENT are significantly larger than
the LO contribution. This reflects the fact that the contribution
from forward jets in the LO scenario is suppressed by kinematics. For
small $x_{Bj}$ the NLO contribution is an order of magnitude larger
than the LO contribution. The NLO contribution opens up the phase
space for forward jets and improves the description of the data
considerably. However, the NLO di-jet predictions are still a factor
of 2 below the data at low $\xbj$. The somewhat improved agreement at
higher $x_{Bj}$ can be understood from the fact that  the range in
the longitudinal momentum fraction which is available for higher
order emissions decreases.

From Fig.~\ref{xfj_djcorr}b it is seen that the CCFM model (both set-1
and set-2) predicts a somewhat harder $x_{Bj}$ distribution, which
results in a comparatively poor description of the data.

Fig.~\ref{xfj_djcorr}c shows that the DGLAP model
with direct photon interactions alone (RG-DIR) gives results similar
to the NLO di-jet calculations and falls below the data, particularly
in the low $x_{Bj}$ region. The description of the data by the
DGLAP model is significantly improved if contributions from resolved
virtual photon interactions are included (RG-DIR+RES). However, there
is still a discrepancy in the lowest  $x_{Bj}$-bin, where a possible
BFKL signal would be expected to show up most prominently. The CDM
model, which gives emissions that are non-ordered in transverse
momentum, shows a behaviour similar to the RG DIR+RES model.
Analytic calculations where resolved photon contributions
are included to NLO order~\cite{Kramer:1999jr} again give 
similar agreement with the data as the RG DIR+RES model~\cite{Jung:1998fu}. 

\subsection{Triple Differential Cross Sections}
\label{sec:triple} 
\begin{figure}[htb!]
  \begin{center}
    \vspace*{1mm}
    \vspace*{1cm}
     \epsfig{figure=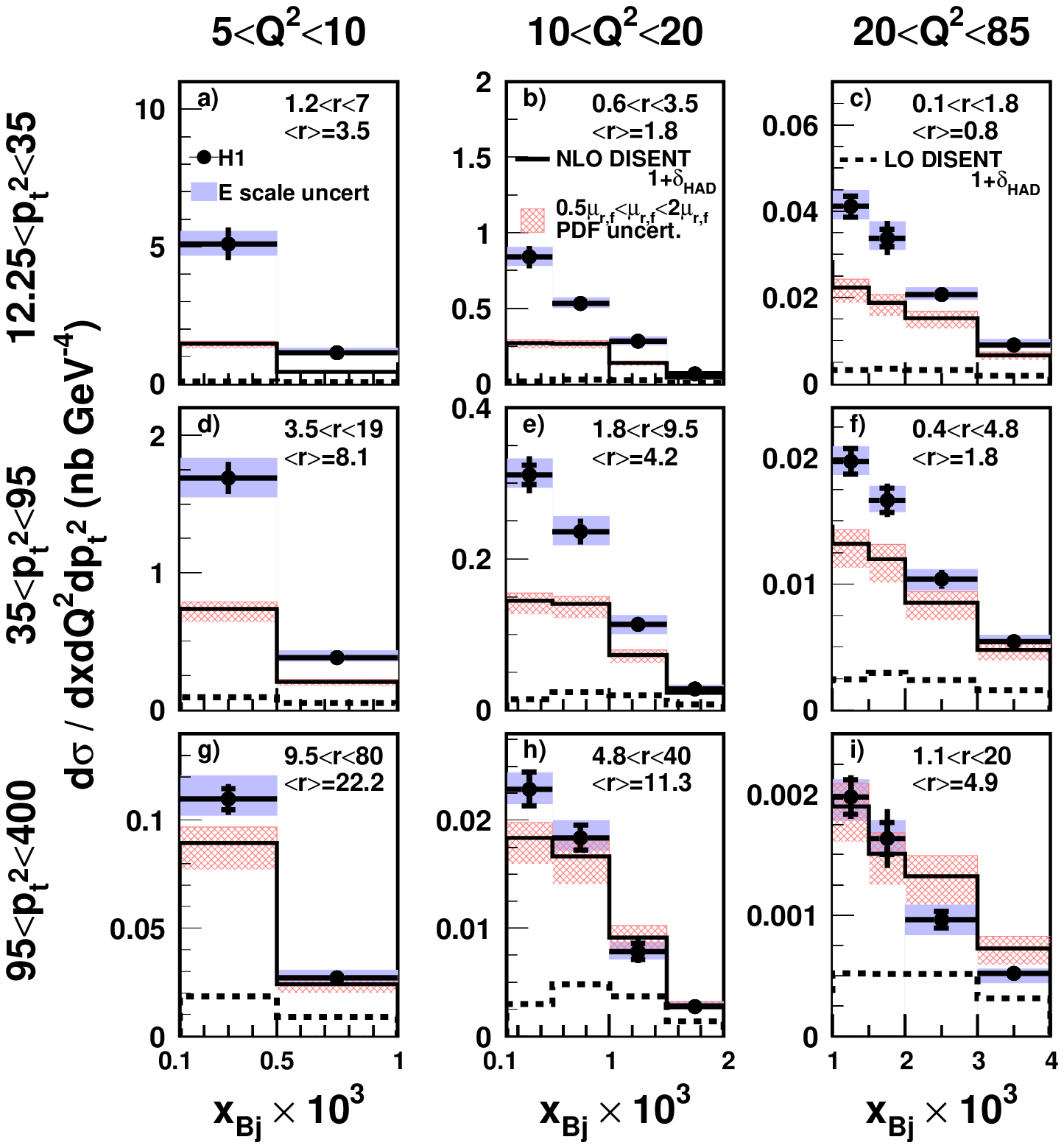,width=15cm,height=17.5cm}
       \caption{{\it
    The hadron level triple differential cross section for forward jet production as a
function of $\xbj$, in bins of $Q^2$ (GeV$^2$) and $p_{t,jet}^2$ (GeV$^2$). The data
are compared to the prediction of NLO (full line) and LO (dashed
line) calculations from DISENT. Both calculations are corrected
for hadronisation effects. The band around the data points
illustrates the error due to the uncertainties in the calorimetric energy 
scales. The band around the NLO calculations illustrates
the theoretical uncertainties in the calculations.  In each bin
the range in and the average value of $r=p_{t,jet}^2/Q^2$ is
shown.
    \label{xp2q2hadnlo}}}
  \end{center}
\end{figure}

\begin{figure}[htb]
  \begin{center}
    \vspace*{1mm}
    \vspace*{1cm}
    \epsfig{figure=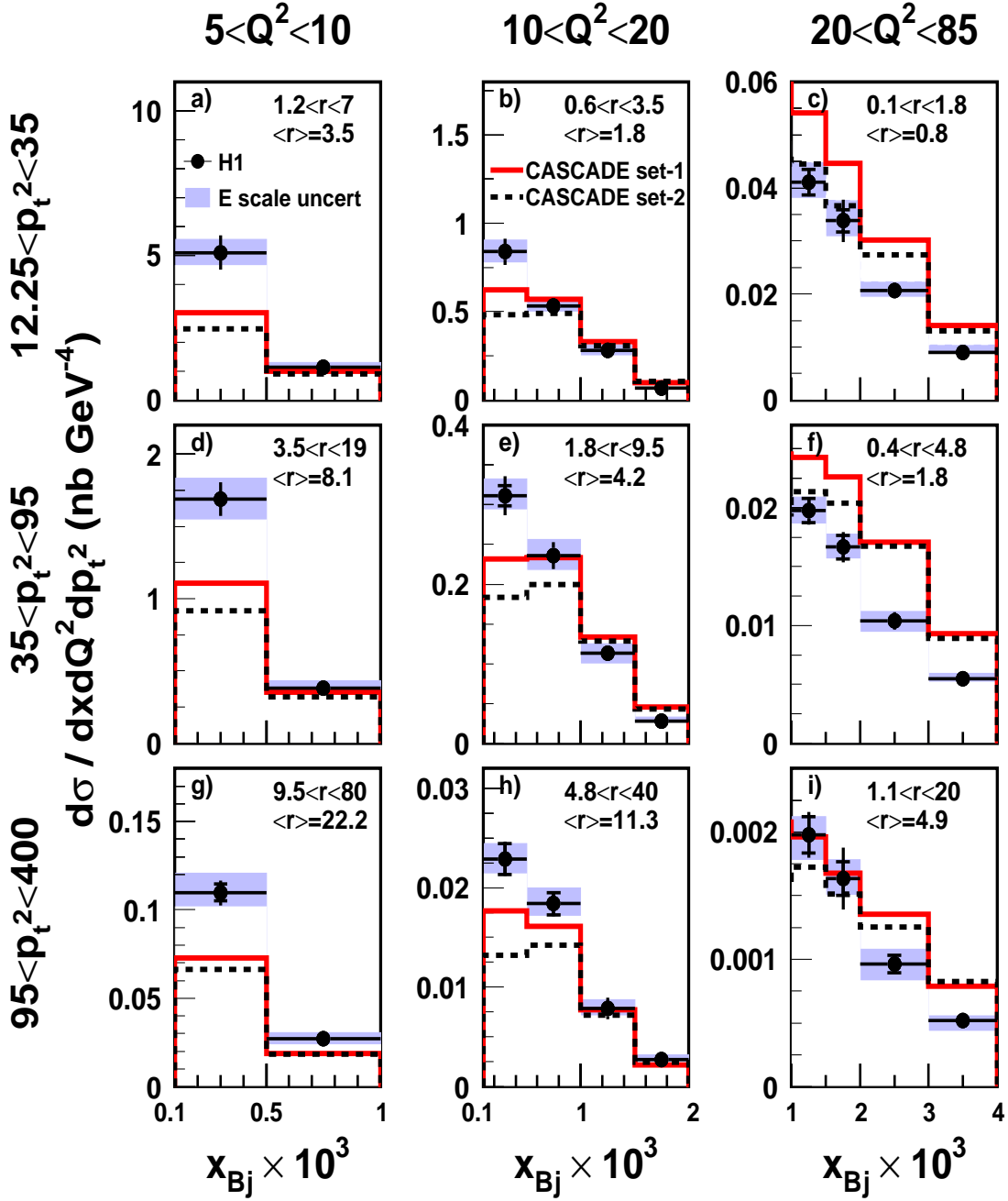,width=15cm,height=17.5cm}
        \caption{{\it
    The hadron level triple differential cross section for forward jet production as a
function of $\xbj$, in bins of $Q^2$ (GeV$^2$) and $p_{t,jet}^2$ (GeV$^2$). The data are
compared to the predictions of CASCADE. The band around the data points
illustrates the error due to the uncertainties in the calorimetric energy 
scales. In each bin the range
in and the average value of $r=p_{t,jet}^2/Q^2$ is shown.
    \label{xp2q2hadcasc}}}
  \end{center}
\end{figure}

\begin{figure}[htb!]
  \begin{center}
    \vspace*{1mm}
    \vspace*{1cm}
    \epsfig{figure=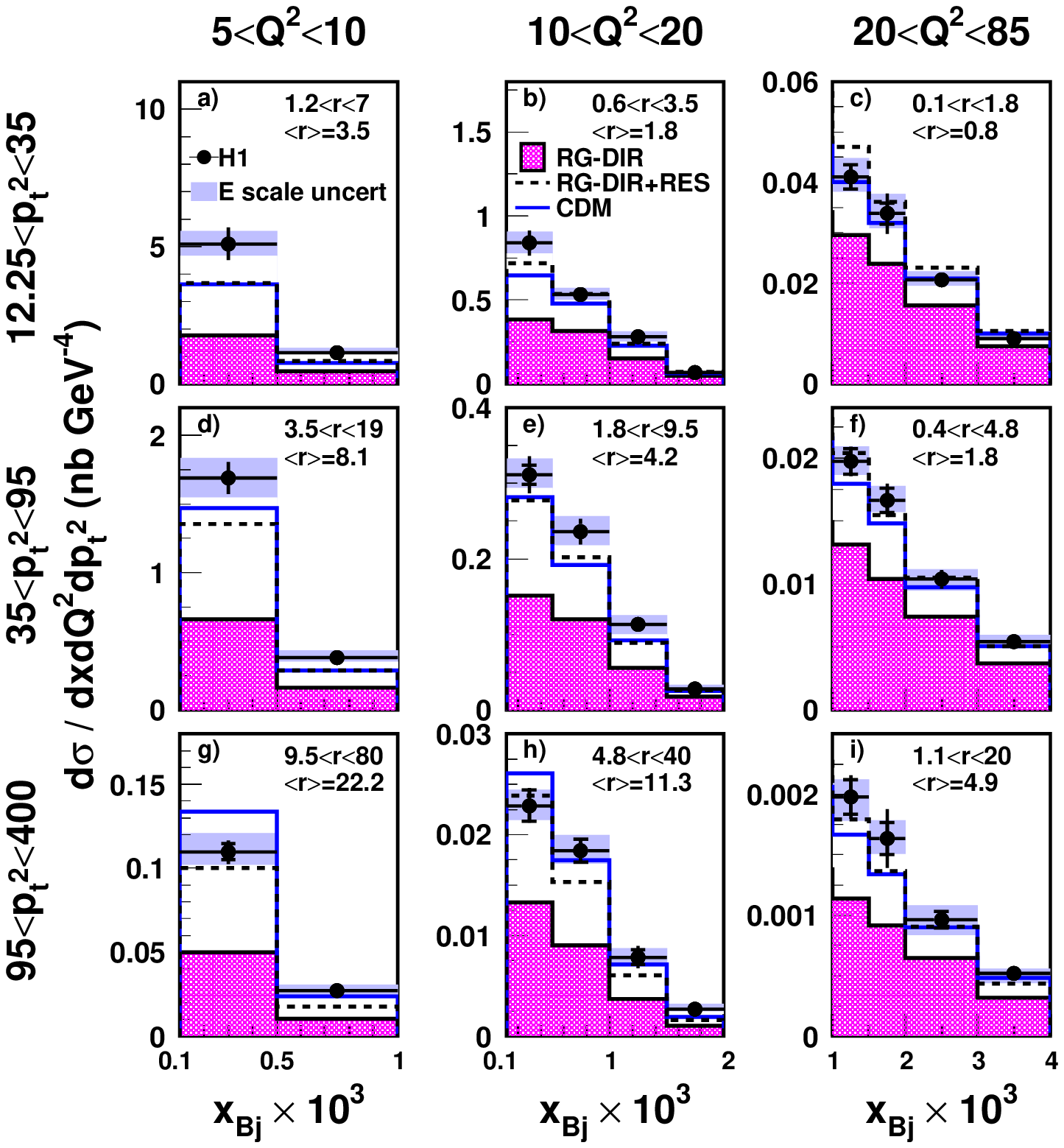,width=15cm,height=17.5cm}
        \caption{{\it
    The hadron level triple differential cross section for forward
jet production as a function of $\xbj$, in bins of $Q^2$ (GeV$^2$)
and $p_{t,jet}^2$ (GeV$^2$). The data are compared to the prediction
of RAPGAP DIR, RAPGAP DIR+RES and CDM. The band around the data points
illustrates the error due to the uncertainties in the calorimetric energy 
scales. In each bin the range in and the average
value of $r=p_{t,jet}^2/Q^2$ is shown.
        \label{xp2q2had}}}
  \end{center}
\end{figure}

In this section data are presented as triple
differential forward jet cross sections. The total forward jet
event sample is subdivided into bins of $Q^2$ and  $p_{t,jet}^2$.
The triple differential cross section
$d\sigma/dx_{Bj}dQ^2dp_{t,jet}^2$ versus $x_{Bj}$ is shown
in Figs.~\ref{xp2q2hadnlo}-\ref{xp2q2had} for three regions in
$Q^2$ and $p_{t,jet}^2$. Fig.~\ref{xp2q2hadnlo} presents the cross
section compared to NLO ($\alpha^2_s$) calculations, including theoretical
errors, represented by error bands.  In Fig.~\ref{xp2q2hadcasc}
and ~\ref{xp2q2had} comparisons to QCD Monte Carlo models are
shown. The same parton density functions and scales are used
as in the measurement of the single differential cross section.
The cross section values are listed in table~\ref{tab:3dif_xsec}.

From Fig.~\ref{xp2q2hadnlo} it can be observed that the NLO
calculations in general undershoot the data but similarly to the
single differential cross section the NLO calculations get closer to
the data at higher $x_{Bj}$ and so too, due to the kinematics, at
higher $Q^2$. The NLO calculations also give a better description of
data for harder forward jets. In the highest $p_{t,jet}^2$-bin the
difference between data and NLO is less than the (large) uncertainty
in the NLO  calculations in several $x_{Bj}$-bins. This is
consistent with the results from a previous measurement on inclusive
jet production~\cite{Adloff:2002ew}. A possible explanation is that
jets with high $p_t$ remove a large fraction of the energy from the
parton ladder, leaving limited energy available for additional
emissions. Thus, the parton ladder is shorter and more like the
NLO configuration. The general trend is that NLO calculations agree
better with data as $x_{Bj}$, $Q^2$ and $p_{t,jet}^2$ increase. For high
$p_{t,jet}$ the phase space for LO starts to open up, which also makes the
NLO prediction more reliable.

The comparisons between data and QCD models are discussed in three
different kinematic regions as specified below. 
These regions are however not strictly
separated, but overlap. In all three regions the CDM and DGLAP
resolved (RG-DIR+RES) models give very similar predictions (see Fig.~\ref{xp2q2had}) indicating
that a breaking of the ordering of the virtuality is necessary to
describe the data.
As already observed in the single differential measurement the CCFM
model predicts a somewhat harder $x_{Bj}$ distribution than seen in
the data. This is true for the full kinematic range and leads to the poor
description of the data as seen in Fig.~\ref{xp2q2hadcasc}.

\noindent{\boldmath $p_{t,jet}^2 \sim Q^2$ $(r \sim 1)$}

\noindent In this region
events with parton emissions ordered in $p_t$ are suppressed, and thus
parton dynamics beyond DGLAP may show up. The
data are best described by the DGLAP resolved model
(RG-DIR+RES) as observed in Fig.~\ref{xp2q2had}b and f.

\noindent {\boldmath {$p_{t,jet}^2 < Q^2$} $(r < 1)$}
  
\noindent The region where
$Q^2$ might become  larger than $p_{t,jet}^2$ is dominated by direct
photon interactions. However, since $r$ can take values up to 1.8 in 
the most DGLAP-like bin (Fig.~\ref{xp2q2had}c), events with
$p_{t,jet}^2$ of the same  order or even greater than $Q^2$ are also
contributing. This gives an admixture of events with emissions
non-ordered in virtuality. This may explain why the
DGLAP direct model (RG-DIR), although closer to the data in this
region than in others, does not give good agreement with the data except
for the highest $x_{Bj}$-bin. The CDM and DGLAP resolved model
(RG-DIR+RES) reproduce the data very well in this region.

\noindent {\boldmath {$p_{t,jet}^2 > Q^2$} $(r > 1)$}

\noindent The kinematic region where
$p_{t,jet}^2$ is larger than $Q^2$ is typical for processes where the
virtual photon is resolved.  As expected the DGLAP resolved model
(RG-DIR+RES) provides a good overall description of the data, again
similar to the CDM model. However, it can be noted that in the
regions where $r$ is the highest and $x_{Bj}$ small,  CDM shows a
tendency to  overshoot the data. DGLAP direct (RG-DIR) gives cross
sections which are too low (see Fig.~\ref{xp2q2had} d, g and h).

\subsection{Events with Reconstructed Di-jets in Addition to the Forward Jet}
Complementary to the analyses reported in
sections~\ref{sec:inclusive} and~\ref{sec:triple}, where the ratio 
$p_{t,jet}^2/Q^2$ has been used to isolate regions where a 
possible BFKL signal is enhanced,
another method is used to control the evolution kinematics in the
analysis reported here. By requiring the reconstruction of the two hardest
jets in the event in addition to the forward jet, different
kinematic regions can be investigated by applying cuts on the jet momenta and their
rapidity separation as described in more detail in
section~\ref{sec:phasespace}.

\begin{figure}[htb]
  \begin{center}
    \vspace*{1mm}
    \vspace*{1cm}
    \epsfig{figure=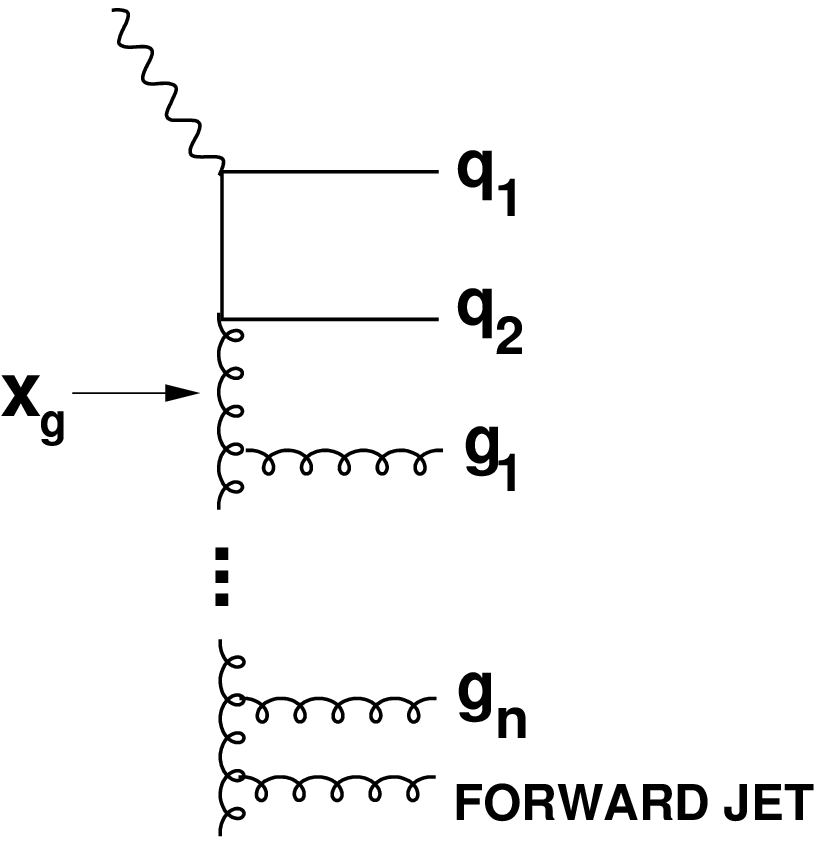,height=5cm}
    \caption{{\it
A schematic diagram of an event giving a forward jet and two
additional hard jets. These may stem from the quarks ($q_1$ and
$q_2$) in the hard scattering vertex or from gluons in
the parton ladder. $x_g$ is the longitudinal momentum fraction
carried by the gluon, connecting to the hard di-jet system
(in this case $q_1$ and $q_2$) .
    \label{fwdjet-dijet}}}
  \end{center}
\end{figure}

In this scenario it is demanded that all jets have transverse momenta
larger than 6 GeV. By applying the same $p_{t,jet}$-cut to all three
jets, evolution with strong $k_{t}$-ordering is not favoured.
Decreasing the $p_{t,jet}$-cut is not possible in this analysis due to detector resolutions.
The jets are ordered in rapidity according to $\eta_{fwdjet} >
\eta_{jet_2} > \eta_{jet_1} > \eta_e$  with $\eta_e$ being the
rapidity of the scattered electron. The
cross section is  measured in two intervals of $\Delta \eta_1 = 
\eta_{jet_2} - \eta_{jet_1} $. If 
the di-jet system originates from the quarks $q_1$ and $q_2$ 
(see Fig.~\ref{fwdjet-dijet}), the phase space for evolution in $x$ between 
the di-jet system and the forward jet is increased by requiring that 
$\Delta\eta_1$ is small and that 
$\Delta\eta_2 = \eta_{fwdjet} - \eta_{jet_2}$ is large. $\Delta\eta_1 < 1$ 
favours small invariant masses of the di-jet system and thereby small values 
of $x_g$ (see Fig.~\ref{fwdjet-dijet}).  With $\Delta\eta_2$ large, 
$x_g$ carries only a small fraction 
of the total propagating momentum, leaving the rest for additional radiation.  
It should be kept in mind, however, that only the forward jet is explicitely 
restricted in rapidity space, by the demand that it has to be close to the 
proton axis.  The directions of the other jets are related to the forward 
jet through the $\Delta\eta$ requirements.  When $\Delta\eta_2$ is small, it is 
therefore possible that one or both of the additional jets originate 
from gluon radiation close in rapidity space to the forward jet. 
With $\Delta\eta_1$ large, BFKL-like evolution may then occur between the 
two jets from the di-jet system, or, with both $\Delta\eta_1$ and $\Delta\eta_2$ small, even between 
the di-jet system and the hard scattering vertex. By studying the cross 
section for different $\Delta\eta$ values one can test theory and models 
for event topologies where the $k_t$ ordering is broken at varying 
locations along the evolution chain.

\begin{figure}[htb]
  \begin{center}
    \vspace*{1mm}
    \vspace*{1cm}
\epsfig{figure=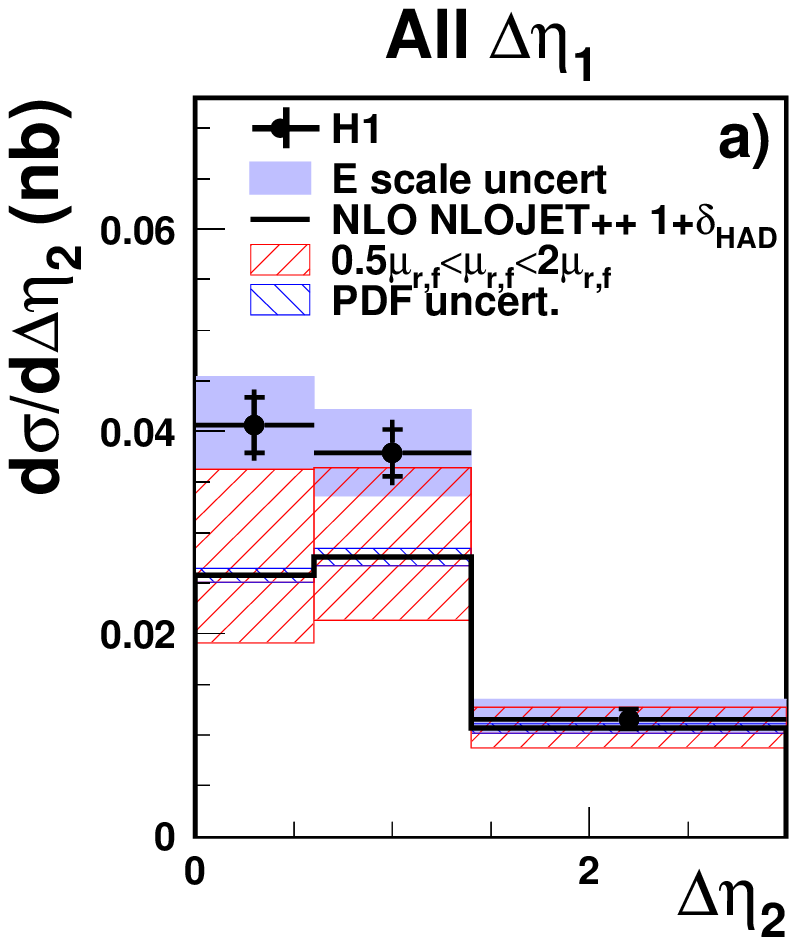,width=4.8cm}
\epsfig{figure=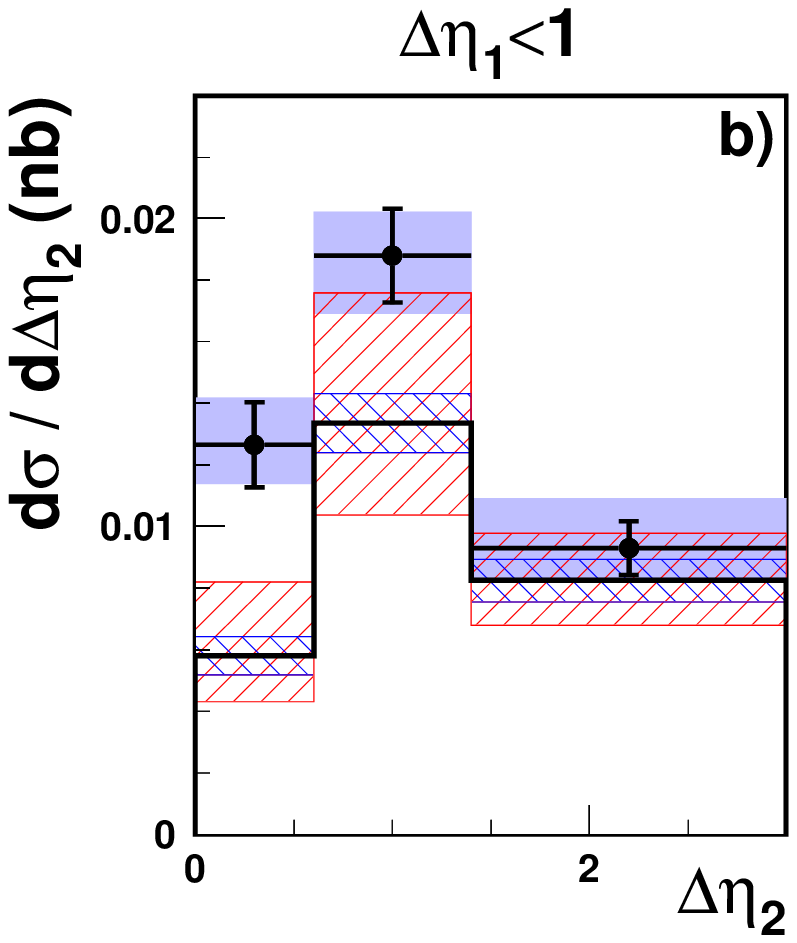,width=4.8cm}
\epsfig{figure=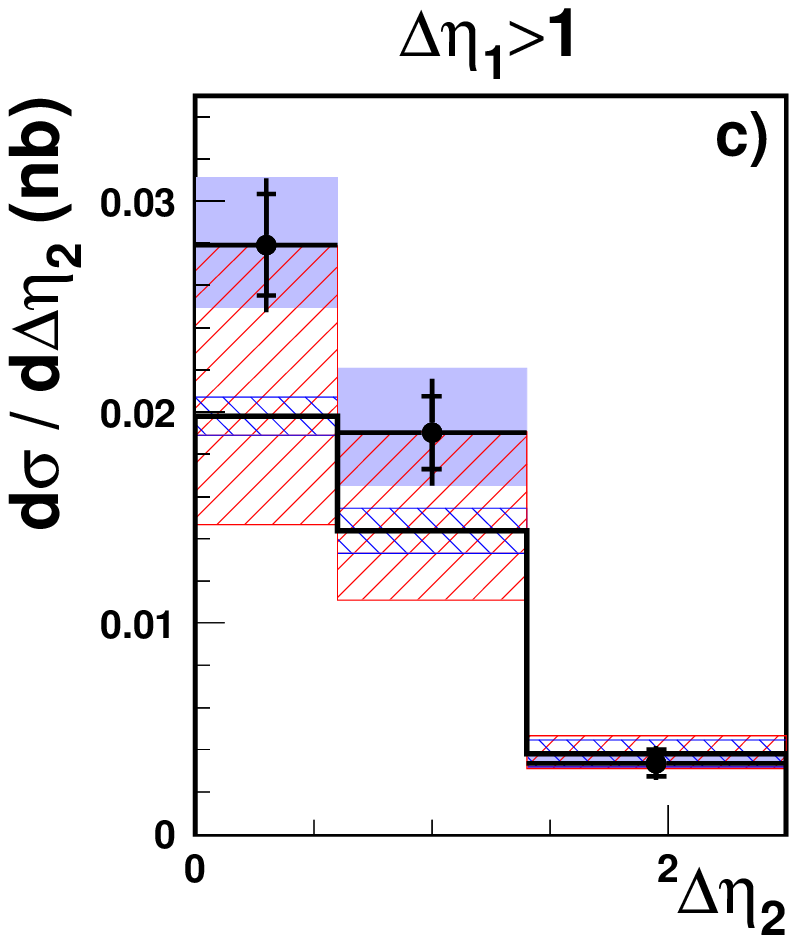,width=4.8cm}
                \caption{{\it
    The cross section for events with a reconstructed high transverse momentum
     di-jet system and a forward
jet as a function of the rapidity separation between the forward jet and the most
forward-going additional jet, $\Delta\eta_2$. Results are shown for the full sample and
for two ranges of the separation between the two additional jets, $\Delta\eta_1<1$ and $\Delta\eta_1>1$.
The data are compared to the predictions of a three-jet NLO calculations from NLOJET++ $(1+\delta_{\textrm{HAD}})$.
The band around the data points
illustrates the error due to the uncertainties in the calorimetric energy 
scales. The band around the NLO
calculations illustrates the theoretical uncertainties in the
calculations.
    \label{2+fwdnlo}}}
  \end{center}
\end{figure}

\begin{figure}[htb]
  \begin{center}
    \vspace*{1mm}
    \vspace*{1cm}
    \epsfig{figure=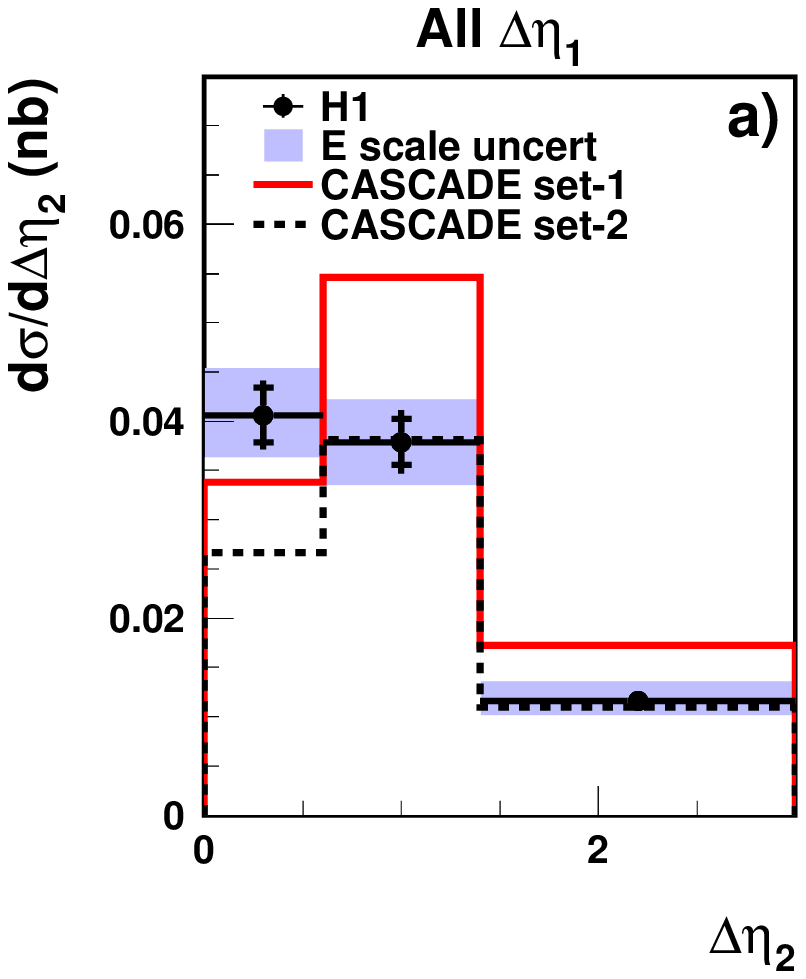,width=4.8cm}
    \epsfig{figure=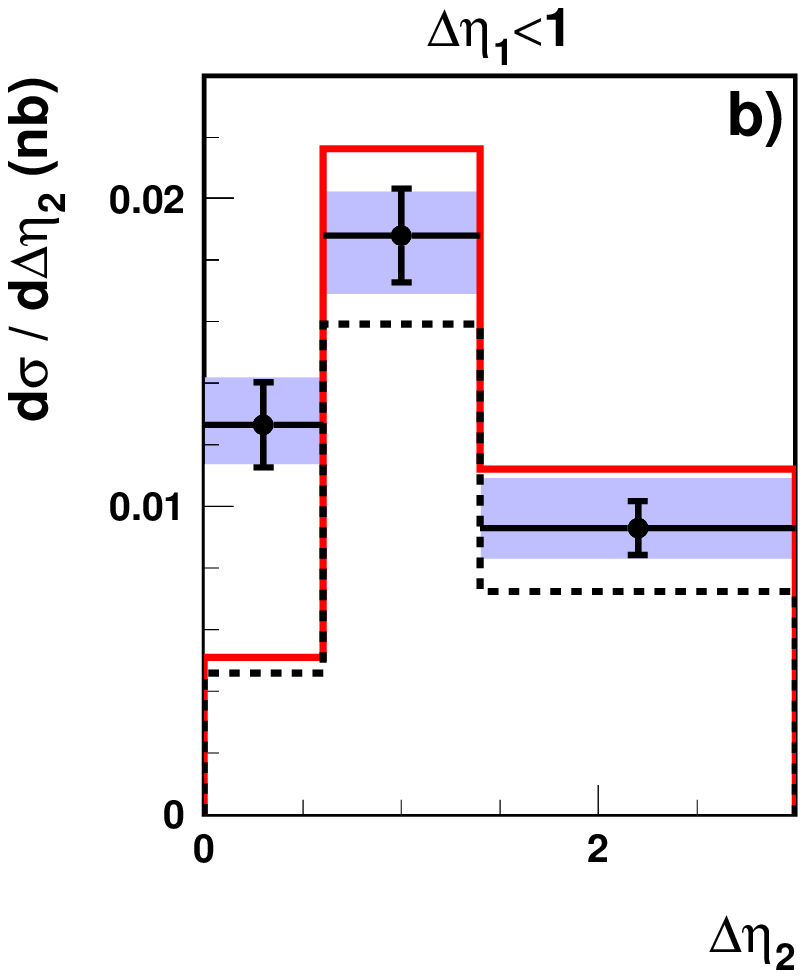,width=4.8cm}
    \epsfig{figure=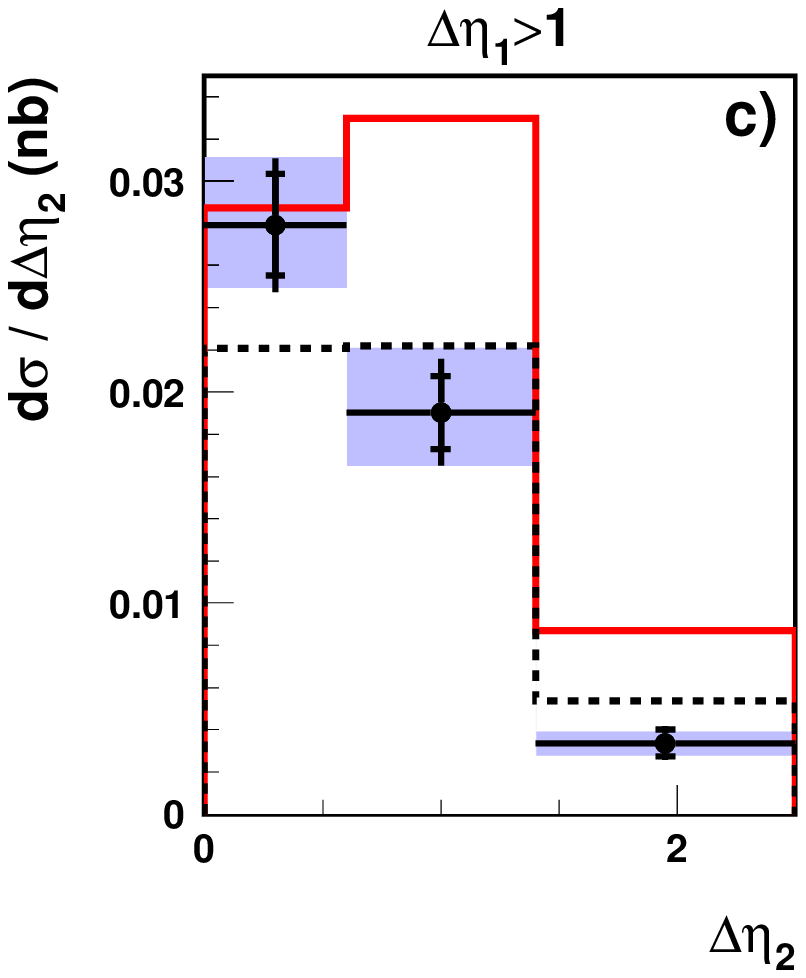,width=4.8cm}
                \caption{{\it
    The cross section for events with a reconstructed high transverse momentum
     di-jet system and a forward
jet as a function of the rapidity separation between the forward jet and the most
forward-going additional jet, $\Delta\eta_2$. Results are shown for the full sample and
for two ranges of the separation between the two additional jets, $\Delta\eta_1<1$ and $\Delta\eta_1>1$.
The data are compared to the predictions of CASCADE. The band around the data points
illustrates the error due to the uncertainties in the calorimetric energy 
scales.
    \label{2+fwdcasc}}}
  \end{center}
\end{figure}

\begin{figure}[htb]
  \begin{center}
    \vspace*{1mm}
    \vspace*{1cm}
    \epsfig{figure=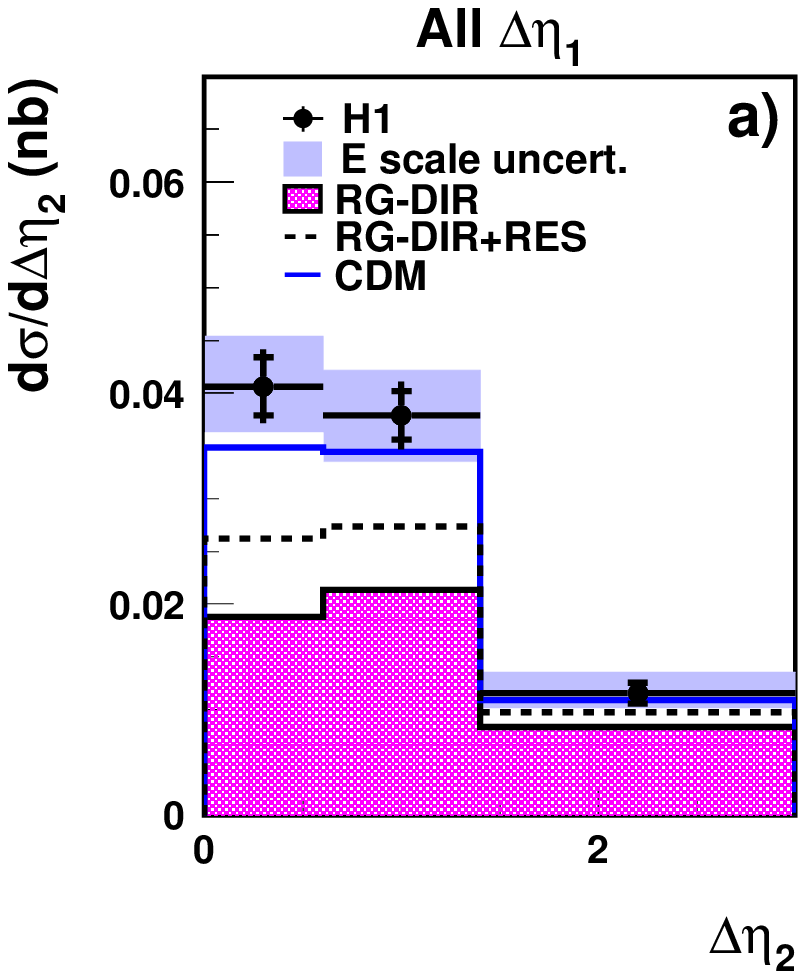,width=4.8cm}
    \epsfig{figure=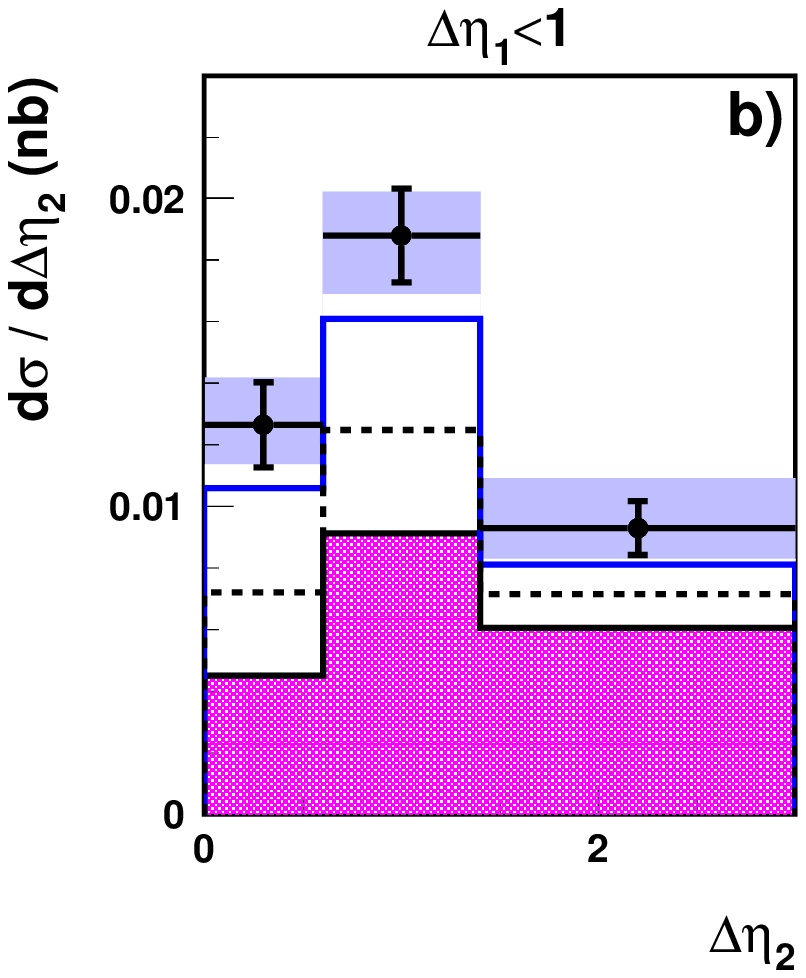,width=4.8cm}
    \epsfig{figure=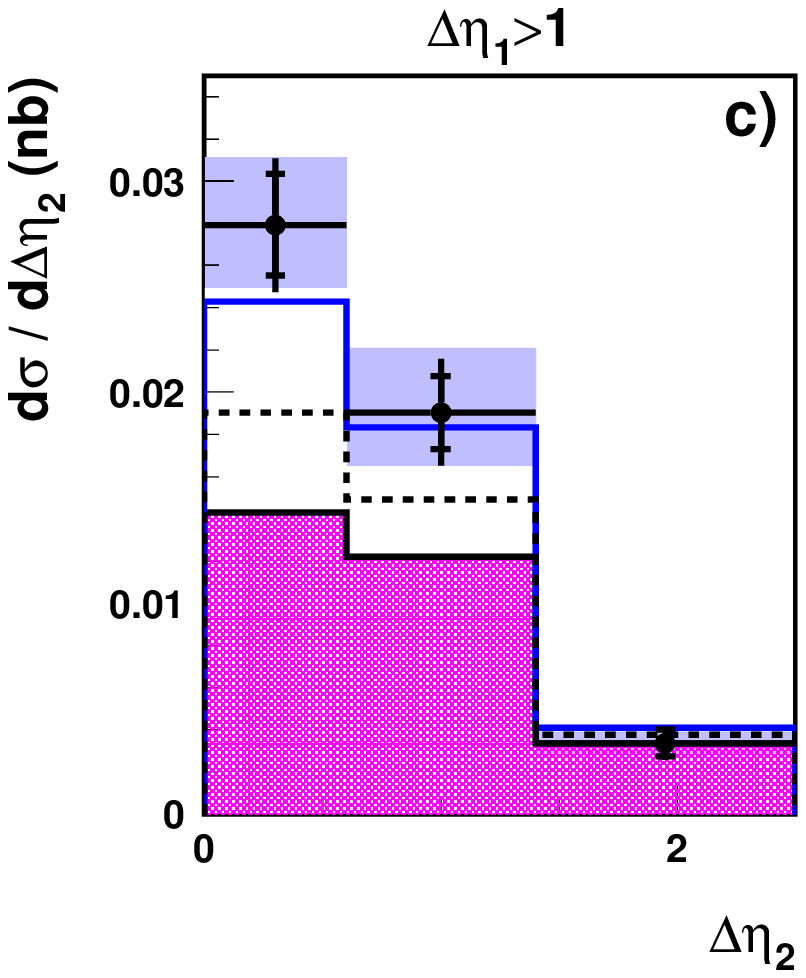,width=4.8cm}
        \caption{{\it The cross section for events with a
reconstructed high transverse momentum di-jet system and a
forward jet as a function of the rapidity separation between the
forward jet and the most forward-going additional jet, $\Delta\eta_2$.
Results are shown for the full sample and for two ranges of the
separation between the two additional jets, $\Delta\eta_1<1$ and
$\Delta\eta_1>1$. The data are compared to the predictions of RAPGAP
DIR, RAPGAP DIR+RES and CDM. The band around the data points
illustrates the error due to the uncertainties in the calorimetric energy 
scales.\label{2+fwdqcd}}}
  \end{center}
\end{figure}

The cross sections for events containing a di-jet system in addition
to the forward jet are presented as a function of  $\Delta \eta_2$ 
in Figs.~\ref{2+fwdnlo}-\ref{2+fwdqcd}
for all `2+forward jet' events , and for the requirements $\Delta
\eta_1<1$  and $\Delta \eta_1>1$, respectively. 
The measured cross sections are given in table~\ref{tab:xsec2b}.
For the $\Delta
\eta_1<1$ region the cross section falls at low $\Delta\eta_2$ since
the phase space becomes smaller when the 3 jets are forced to be
close together. Fig.~\ref{2+fwdnlo} gives a comparison of data to NLO
($\alpha^3_s$) predictions with theoretical error contributions
included as bands. In Figs.~\ref{2+fwdcasc} and~\ref{2+fwdqcd}
comparisons to QCD models are presented.

In this investigation the same settings of the QCD models are used as
in sections~\ref{sec:inclusive} and ~\ref{sec:triple}, while the NLO three-jet
cross sections are calculated using NLOJET++.

From Fig.~\ref{2+fwdnlo} it is observed that NLOJET++ gives good 
agreement with the data if the two additional hard jets are emitted
in the central region ($\Delta\eta_2$ large). It is  interesting to
note that a fixed order calculation ($\alpha_s^3$),  including the
$\log (1/x)$-term to the first order in $\alpha_s$, is  able to
describe these data well. However, the more the additional  hard jets
are shifted to the forward region ($\Delta\eta_2$ small), the less
well are the data described by NLOJET++. This can be understood from
the fact that the more forward the additional jets go, the higher 
the probability is that one of them, or even both, do not actually
originate from quarks but from additional radiated gluons.
For gluon induced processes, which dominate at small $x$, 
NLOJET++ calculates the NLO contribution to final states  containing
one gluon jet and two jets from the di-quarks, i.e. it accounts for
the emission  of one gluon in addition to the three jets.
Thus, events where two of the three selected jets originate from gluons
are produced by NLOJET++ only in the real emission corrections to the
three-jet final state, which effectively means that these kinematic
configurations are only produced to leading order ($\alpha_s^3$). The
most extreme case, where all three reconstructed jets are produced by
gluons, is not considered by NLOJET++.
This results in a depletion of the
theoretical cross section in the small $\Delta\eta_2$ region, which
is more pronounced when $\Delta\eta_1$ is also small, i.e. when all
three jets are in the forward region. Consequently a significant
deviation between data and NLOJET++ can be observed for such events
(see the lowest bin in Fig.~\ref{2+fwdnlo}b).
Accounting for still higher orders in $\alpha_s$ might improve the
description of the data in this domain since virtual corrections to the
production of two gluons could increase the cross section for such final
states, and additional gluon emissions would enhance the probability that
one of the soft radiated gluons produces a jet that fulfills the
transverse momentum requirement applied in this analysis.  

For the `2+forward jet' sample CCFM is not
describing well the shape of the $\eta$-distributions (Fig.~\ref{2+fwdcasc}a,
b and c).

As explained above, evolution with strong $k_t$-ordering is 
disfavoured in this study. Radiation that is non-ordered in $k_t$ 
may occur at different locations along the evolution chain, 
depending on the values of $\Delta\eta_1$ and $\Delta\eta_2$.  
As can be seen from Fig. 10, the colour dipole model gives 
good agreement in all cases, whereas the DGLAP models 
give cross sections that are too low except when both $\Delta\eta_1$ 
and $\Delta\eta_2$ are large. For this last topology all models 
and the NLO calculation agree with the data, indicating that  
the available phase space is exhausted and that little freedom 
is left for dynamical variations.

If one or both jets from the di-jet system are produced by gluon 
radiation, which is increasingly probable the more forward these 
jets go, it necessarily means that the $k_t$ ordering is broken.  In
this  context it is noteworthy that CDM provides the best
description  of the data while the other models, including the
DGLAP-resolved  model, fail in most of the bins.  The `2+forward jet'
sample  differentiates CDM and the DGLAP-resolved model, in contrast
to  the more inclusive samples where CDM and RG-DIR+RES give the
same  predictions.  The conclusion is that additional breaking of
the  $k_t$ ordering is needed compared to what is included in the 
resolved photon model.

\section{Summary} An investigation of DIS events containing a jet in
the forward direction is presented. Various constraints are
applied,  which suppress contributions to the parton evolution
described by the DGLAP equations and enhance the sensitivity to
other  parton dynamics. Several observables involving forward jet
events are studied and compared to the predictions of NLO
calculations and QCD models.

Leading order ($\alpha_s$) calculations of the single differential forward
jet cross section, $d\sigma / dx_{Bj}$, are well below the measurements, 
which is expected since forward jet production is kinematically suppressed
in LO. NLO di-jet calculations improve the description of the data but remain 
too low at small values of $x_{Bj}$. This is also the case for predictions
based on the DGLAP direct model. The DGLAP resolved
photon model (RG-DIR+RES) and the colour dipole model (CDM) come closest
to the data.

The total forward jet sample is subdivided into bins of $Q^2$ and
$p_{t,jet}^2$ such that kinematic regions are defined in which
the effects of different evolution dynamics are enhanced. 
In the most
DGLAP enhanced region, ($Q^2 \gg p_{t,jet}^2$), and in the region where
contributions from resolved processes are expected to become
important ($p_{t,jet}^2 \gg Q^2$), the measured triple 
differential forward jet cross sections are well described by the CDM
and the DGLAP resolved model (RG-DIR+RES). In the BFKL region ($Q^2
\sim p_{t,jet}^2$) the CDM and DGLAP resolved model (RG-DIR+RES)
again reproduce the data best. A general observeration is that the
DGLAP resolved model and CDM tend to fall below the data at low
$\xbj$, $Q^2$ and $p^2_t$. The cross sections predicted by the DGLAP
direct model (RG-DIR) are consistently too low in all regions and
especially at low $x_{Bj}$. 

The NLO di-jet calculations from DISENT describe the data for the largest
values of $\xbj$ at high values of $Q^2$ and $p^2_t$, but fail for
low values of these variables.

The measured cross section for events with a reconstructed di-jet system in
addition to the forward jet are in good agreement with the
predictions of NLOJET++~ if the
additional jets are emitted in the central region. As expected, deviations 
are observed for other jet topologies.
The data are best described by the CDM. The
DGLAP resolved model (RG-DIR+RES) is below the data as is, to an 
even greater extent, the DGLAP direct model (RG-DIR). 
This result gives the first evidence for parton dynamics in
which there is additional breaking of the $k_t$-ordering compared to
that provided by the resolved photon model. 

The CCFM model, as implemented in
CASCADE, with two different parametrisations of the unintegrated
gluon density, fails to describe the shape of both the single and triple
differential cross sections, as well as the `2+forward jet' cross section.
This might be caused by the
parametrisation of the unintegrated gluon density and/or
the missing contributions from splittings into quark pairs. 

The observations made here demonstrate that an accurate description
of the radiation pattern at small $x_{Bj}$ requires the introduction
of terms beyond those included in the DGLAP direct approximation
(RG-DIR). Higher order parton emissions with breaking of the
transverse momentum ordering contribute noticeably to the cross
section. Calculations which include such processes, such as CDM and
the resolved photon model, provide a better description of the data.
The similar behaviour of CDM and the DGLAP resolved model
(RG-DIR+RES), which describe the data best, indicates that the inclusive
forward jet measurements do not give a significant
separation of the models. However, in the more exclusive measurement 
of `2+forward jet' events a clear differentiation of the models is 
obtained since, in contrast to CDM, the DGLAP resolved model (RG-DIR+RES) 
fails to describe the data.

\linespread{1.35}
\begin{table}[h]
\begin{center}
\begin{tabular}{|c||c|c|c|c|}
\hline
 $\xbj$ & $d\sigma/d\xbj$ (nb)& $\Delta_{\textrm{Stat}}$ (nb)& $\Delta_{\textrm{Syst1}}$ (nb)& $\Delta_{\textrm{Syst2}}$ (nb)\\
\hline
\hline
 0.0001-0.0005 & 925 & $\pm$ 17  & \se{110}{100} &  \se{77}{77} \\
\hline
0.0005-0.001 & 541 & $\pm$ 12 &\se{54}{55} & \se{23}{24} \\
\hline
0.001-0.0015 & 264 & $\pm$ 8  & \se{30}{28} & \se{11}{11}  \\
\hline
0.0015-0.002 & 153 & $\pm$ 6  & \se{19}{16} & \se{8}{8}  \\
\hline
0.002-0.003 & 74.5 & $\pm$ 3.0  & \se{10.7}{8.0} & \se{1.9}{1.8}  \\
\hline
0.003-0.004 & 36.7 & $\pm$ 2.0  & \se{2.1}{5.7} & \se{2.4}{2.4}   \\
\hline
\end{tabular}
\end{center}
\linespread{1.0}
\caption{{\it Single differential forward jet cross sections in bins of
$x_{Bj}$. The statistical error ($\Delta_{\textrm{Stat}}$), the error from
the uncertainty of the calorimetric energy scales ($\Delta_{\textrm{Syst1}}$)
and from the other systematic errors ($\Delta_{\textrm{Syst2}}$) are
specified. \label{tblxsecxfj}}}
\end{table}

\begin{table}[h]
\begin{center}
\begin{tabular}{|c|c|c||c|c|c|c|}
\hline
 $Q^2 \atop (\textrm{GeV}^2)$ & $p^2_t \atop (\textrm{GeV}^2)$ & $\xbj$ & $d^3\sigma/dx_{Bj}dQ^2dp^2_t \atop (\textrm{nb GeV}^{-4})$ & $\Delta_{\textrm{Stat}} \atop (\textrm{nb})$ & $\Delta_{\textrm{Syst1}} \atop (\textrm{nb})$ & $\Delta_{\textrm{Syst2}} \atop (\textrm{nb})$\\
\hline
\hline
      & 12.25-35  & 0.0001-0.0005 & 5.10   &  $\pm$ 0.12   & \se{0.46}{0.44} & \se{0.58}{0.59} \\
        				 \cline{4-7}
      &           & 0.0005-0.001 &  1.13   &  $\pm$ 0.05 & \se{0.16}{0.07} &  \se{0.17}{0.17} \\
      \cline{2-7}
 5-10 & 35-95     & 0.0001-0.0005 &  1.70   &  $\pm$ 0.04   &  \se{0.14}{0.14} &  \se{0.11}{0.11} \\
        				 \cline{4-7}
      &           & 0.0005-0.001 &  3.81$ \cdot 10^{-1}$  &  $\pm$ 0.18$ \cdot 10^{-1}$   &  \se{0.51 \cdot 10^{-1}}{0.33 \cdot 10^{-1}} &  \se{0.13 \cdot 10^{-1}}{0.10 \cdot 10^{-1}} \\
      \cline{2-7}
      & 95-400    & 0.0001-0.0005 &  1.11$ \cdot 10^{-1}$   &  $\pm$ 0.05$ \cdot 10^{-1}$   &  \se{0.11 \cdot 10^{-1}}{0.08 \cdot 10^{-1}} &  \se{0.05 \cdot 10^{-1}}{0.05 \cdot 10^{-1}} \\
        				 \cline{4-7}
      &           & 0.0005-0.001 &  2.71$ \cdot 10^{-2}$  &  $\pm$ 0.22$ \cdot 10^{-2}$   & \se{0.35 \cdot 10^{-2}}{0.33 \cdot 10^{-2}} & \se{0.26 \cdot 10^{-2}}{0.26 \cdot 10^{-2}} \\
\hline
\hline
       & 12.25-35 & 0.0001-0.0005 &  8.40$ \cdot 10^{-1}$   &  $\pm$ 0.31$ \cdot 10^{-1}$   &  \se{0.64 \cdot 10^{-1}}{0.62 \cdot 10^{-1}} & \se{0.67 \cdot 10^{-1}}{0.66 \cdot 10^{-1}} \\
        				 \cline{4-7}
       &          & 0.0005-0.001 &  5.31$ \cdot 10^{-1}$   &  $\pm$ 0.24$ \cdot 10^{-1}$   &  \se{0.39 \cdot 10^{-1}}{0.34 \cdot 10^{-1}}  &  \se{0.22 \cdot 10^{-1}}{0.22 \cdot 10^{-1}} \\
        				 \cline{4-7}
       &          & 0.001-0.0015 & 2.81$ \cdot 10^{-1}$   &  $\pm$ 0.16$ \cdot 10^{-1}$  &  \se{0.32 \cdot 10^{-1}}{0.29 \cdot 10^{-1}} &  \se{0.36 \cdot 10^{-1}}{0.37 \cdot 10^{-1}}  \\
        				 \cline{4-7}
       &          & 0.0015-0.002 &  6.67$ \cdot 10^{-2}$   &  $\pm$ 0.73$ \cdot 10^{-2}$   & \se{0.38 \cdot 10^{-2}}{0.65 \cdot 10^{-2}} &  \se{0.09 \cdot 10^{-2}}{0.08 \cdot 10^{-2}}   \\
       \cline{2-7}
       & 35-95    & 0.0001-0.0005 &  3.11$ \cdot 10^{-1}$   &  $\pm$ 0.13$ \cdot 10^{-1}$   & \se{0.21 \cdot 10^{-1}}{0.17 \cdot 10^{-1}} &  \se{0.21 \cdot 10^{-1}}{0.21 \cdot 10^{-1}} \\
        				 \cline{4-7}
 10-20 &          & 0.0005-0.001  & 2.36$ \cdot 10^{-1}$   &  $\pm$ 0.09$ \cdot 10^{-1}$   & \se{0.20 \cdot 10^{-1}}{0.18 \cdot 10^{-1}} &  \se{0.14 \cdot 10^{-1}}{0.15 \cdot 10^{-1}}  \\
         				 \cline{4-7}
      &           & 0.001-0.0015  & 1.13$ \cdot 10^{-1}$     &  $\pm$ 0.06$ \cdot 10^{-1}$   & \se{0.12 \cdot 10^{-1}}{0.13 \cdot 10^{-1}} & \se{0.03 \cdot 10^{-1}}{0.03 \cdot 10^{-1}}  \\
        				 \cline{4-7}
       &          & 0.0015-0.002 &  2.81$ \cdot 10^{-2}$ &  $\pm$ 0.33$ \cdot 10^{-2}$   & \se{0.50 \cdot 10^{-2}}{0.19 \cdot 10^{-2}} &  \se{0.27 \cdot 10^{-2}}{0.22 \cdot 10^{-2}} \\
       \cline{2-7}
       & 95-400   & 0.0001-0.0005 &  2.29$ \cdot 10^{-2}$   &  $\pm$ 0.16$ \cdot 10^{-2}$   & \se{0.15 \cdot 10^{-2}}{0.15 \cdot 10^{-2}} &  \se{0.08 \cdot 10^{-2}}{0.07 \cdot 10^{-2}} \\
        				 \cline{4-7}
       &          & 0.0005-0.001 &  1.84$ \cdot 10^{-2}$   &  $\pm$ 0.11$ \cdot 10^{-2}$   & \se{0.16 \cdot 10^{-2}}{0.13 \cdot 10^{-2}} &  \se{0.04 \cdot 10^{-2}}{0.05 \cdot 10^{-2}} \\
        				 \cline{4-7}
       &          & 0.001-0.0015 &  7.83$ \cdot 10^{-3}$   &  $\pm$ 0.74$ \cdot 10^{-3}$   & \se{0.87 \cdot 10^{-3}}{0.75 \cdot 10^{-3}} &  \se{0.83 \cdot 10^{-3}}{0.79 \cdot 10^{-3}} \\
        				 \cline{4-7}
       &          & 0.0015-0.002 &  2.70$ \cdot 10^{-3}$   &  $\pm$ 0.45$ \cdot 10^{-3}$   & \se{0.46 \cdot 10^{-3}}{0.27 \cdot 10^{-3}} &  \se{0.35 \cdot 10^{-3}}{0.39 \cdot 10^{-3}} \\
\hline
\hline
       & 12.25-35 & 0.001-0.0015 &  4.11$ \cdot 10^{-2}$   &  $\pm$ 0.24$ \cdot 10^{-2}$   &  \se{0.37 \cdot 10^{-2}}{0.30 \cdot 10^{-2}} & \se{0.12 \cdot 10^{-2}}{0.12 \cdot 10^{-2}} \\
       				 \cline{4-7}
       &          & 0.0015-0.002 & 3.38$ \cdot 10^{-2}$   &  $\pm$ 0.21$ \cdot 10^{-2}$   & \se{0.38 \cdot 10^{-2}}{0.29 \cdot 10^{-2}} & \se{0.34 \cdot 10^{-2}}{0.34 \cdot 10^{-2}} \\
       				 \cline{4-7}
       &          & 0.002-0.003  & 2.07$ \cdot 10^{-2}$   &  $\pm$ 0.12$ \cdot 10^{-2}$   &  \se{0.16 \cdot 10^{-2}}{0.13 \cdot 10^{-2}} & \se{0.06 \cdot 10^{-2}}{0.07 \cdot 10^{-2}} \\
        				 \cline{4-7}
      &          & 0.003-0.004  & 9.03$ \cdot 10^{-3}$   &  $\pm$ 0.79$ \cdot 10^{-3}$   & \se{1.37 \cdot 10^{-3}}{0.12 \cdot 10^{-3}} & \se{0.44 \cdot 10^{-3}}{0.44 \cdot 10^{-3}} \\
       \cline{2-7}
       & 35-95    & 0.001-0.0015  & 1.97$ \cdot 10^{-2}$   &  $\pm$ 0.10$ \cdot 10^{-2}$  &  \se{0.11 \cdot 10^{-2}}{0.11 \cdot 10^{-2}} & \se{0.06 \cdot 10^{-2}}{0.05 \cdot 10^{-2}} \\
       				 \cline{4-7}
 20-85 &          & 0.0015-0.002 & 1.67$ \cdot 10^{-2}$   &  $\pm$ 0.10$ \cdot 10^{-2}$  & \se{0.11 \cdot 10^{-2}}{0.10 \cdot 10^{-2}} & \se{0.09 \cdot 10^{-2}}{0.09 \cdot 10^{-2}} \\
         				 \cline{4-7}
     &          & 0.002-0.003  & 1.04$ \cdot 10^{-2}$    &  $\pm$ 0.06$ \cdot 10^{-2}$  & \se{0.08 \cdot 10^{-2}}{0.10 \cdot 10^{-2}} & \se{0.05 \cdot 10^{-2}}{0.05 \cdot 10^{-2}} \\
        				 \cline{4-7}
      &          & 0.003-0.004  & 5.45$ \cdot 10^{-3}$   &  $\pm$ 0.39$ \cdot 10^{-3}$  & \se{0.46 \cdot 10^{-3}}{0.24 \cdot 10^{-3}} & \se{0.46 \cdot 10^{-3}}{0.46 \cdot 10^{-3}} \\
       \cline{2-7}
       & 95-400   & 0.001-0.0015  & 1.98$ \cdot 10^{-3}$   &  $\pm$ 0.14$ \cdot 10^{-3}$   & \se{0.15 \cdot 10^{-3}}{0.20 \cdot 10^{-3}} & \se{0.11 \cdot 10^{-3}}{0.11 \cdot 10^{-3}} \\
        				 \cline{4-7}
       &          & 0.0015-0.002 & 1.63$ \cdot 10^{-3}$   &  $\pm$ 0.13$ \cdot 10^{-3}$   & \se{0.15 \cdot 10^{-3}}{0.13 \cdot 10^{-3}} & \se{0.20 \cdot 10^{-3}}{0.20 \cdot 10^{-3}} \\
        				 \cline{4-7}
       &          & 0.002-0.003  & 9.64$ \cdot 10^{-4}$   &  $\pm$ 0.70$ \cdot 10^{-4}$   & \se{1.15 \cdot 10^{-4}}{1.35 \cdot 10^{-4}} & \se{0.07 \cdot 10^{-4}}{0.07 \cdot 10^{-4}} \\
        				 \cline{4-7}
       &          & 0.003-0.004  &5.17$ \cdot 10^{-4}$   &  $\pm$ 0.49$ \cdot 10^{-4}$   & \se{0.41 \cdot 10^{-4}}{0.81 \cdot 10^{-4}} & \se{0.06 \cdot 10^{-4}}{0.03 \cdot 10^{-4}} \\
\hline
\end{tabular}
\end{center}
\linespread{1.0}
\caption{ {\it Triple differential cross sections in bins of $Q^2$,
$p^2_t$ and $\xbj$. The statistical error ($\Delta_{\textrm{Stat}}$), the error from
the uncertainty of the calorimetric energy scales ($\Delta_{\textrm{Syst1}}$)
and from the other systematic errors ($\Delta_{\textrm{Syst2}}$) are
specified.  \label{tab:3dif_xsec}}}
\end{table}

\begin{table}[h]
\begin{center}
\begin{tabular}{|c|c||c|c|c|c|}
\hline
 $\Delta \eta_1$ & $\Delta \eta_2$ & $d\sigma/d\Delta \eta_2$ (pb)& $\Delta_{\textrm{Stat}}$ (pb)& $\Delta_{\textrm{Syst1}}$ (pb)& $\Delta_{\textrm{Syst2}}$ (nb)\\
\hline
 	   		& 0.0 - 0.6  & 40.6 & $\pm 2.7$  & \se{4.8}{4.4} & \se{2.1}{2.2}  \\
       				\cline{2-6}
All  $\Delta \eta_1$      & 0.6 - 1.4  & 37.9 & $\pm 2.2$  & \se{4.3}{4.4} & \se{2.2}{2.2}  \\
       				\cline{2-6}
           		& 1.4 - 3.0  & 11.6 & $\pm 1.0$ & \se{2.0}{1.5} & \se{0.2}{0.2}  \\
\hline 
           		& 0.0 - 0.6  & 12.7 & $\pm 1.3$  & \se{1.5}{1.3} & \se{0.3}{0.4}  \\
       				\cline{2-6}
$\Delta \eta_1 < 1$      & 0.6 - 1.4  & 18.8 & $\pm 1.5$  & \se{1.4}{1.9} & \se{0.4}{0.4}  \\
         		 	\cline{2-6}
           		& 1.4 - 3.0  & 9.3 & $\pm 0.9$ & \se{1.6}{1.0} & \se{0.3}{0.3}  \\
\hline 
            		& 0.0 - 0.6  & 27.9 & $\pm 2.4$  & \se{3.2}{3.0} & \se{2.1}{2.1}  \\
        		 	\cline{2-6}
$\Delta \eta_1 > 1$     & 0.6 - 1.4  & 19.0 & $\pm 1.7$  & \se{3.0}{2.6} & \se{1.8}{1.9}  \\
        			\cline{2-6}
           		& 1.4 - 2.5  & 3.4 & $\pm 0.6$ & \se{0.5}{0.6} & \se{0.5}{0.5}  \\
        			\cline{2-6}
\hline
\end{tabular}
\end{center}
\linespread{1.0}
\caption{ {\it `2+forward jet' cross sections in bins of $\Delta \eta_2$ for all $\Delta \eta_1$, $\Delta \eta_1<1$ and $\Delta \eta_1>1$. 
The statistical error ($\Delta_{\textrm{Stat}}$), the error from
the uncertainty of the calorimetric energy scales ($\Delta_{\textrm{Syst1}}$)
and from the other systematic errors ($\Delta_{\textrm{Syst2}}$) are
specified.  \label{tab:xsec2b}}}
\end{table}

\section{Acknowledgement}
We are grateful to the HERA machine group whose outstanding efforts have made and
continue to make this experiment possible. We thank the engineers and technicians for
their work in constructing and maintaining the H1 detector, our funding agencies for
financial support, the DESY technical staff for continual assistance and the DESY
directorate for hospitality which they extend to the non-DESY members of the
collaboration.
J. Bartels, G. Gustafson and L. L\"onnblad are acknowledged for valuable discussions
concerning the interpretation of the results.

\end{document}